\documentclass[aps,pre,twocolumn,superscriptaddress]{revtex4-2}

\usepackage{latexsym}
\usepackage{amsmath, amsthm, amssymb}
\usepackage{mathrsfs}
\usepackage{epsfig}
\usepackage{graphicx}
\usepackage{dcolumn}
\usepackage{expl3}
\usepackage{float}

\usepackage{tikz}
\usetikzlibrary{trees}

\usepackage{natbib}
\usepackage{hyperref}
\usepackage{placeins}

\allowdisplaybreaks

\begin{document}

\title{A hierarchy index for networks in the brain reveals a complex entangled organizational structure} 

\author{Anand Pathak}
\affiliation{The Institute of Mathematical Sciences, CIT Campus, Taramani, Chennai 600113, India}
\affiliation{Homi Bhabha National Institute, Anushaktinagar, Mumbai 400 094, India}

\author{Shakti~N.~Menon}
\affiliation{The Institute of Mathematical Sciences, CIT Campus, Taramani, Chennai 600113, India}

\author{Sitabhra~Sinha}
\affiliation{The Institute of Mathematical Sciences, CIT Campus, Taramani, Chennai 600113, India}
\affiliation{Homi Bhabha National Institute, Anushaktinagar, Mumbai 400 094, India}

\date{\today}



\begin{abstract}
Networks involved in information processing often have their nodes arranged hierarchically,
with the majority of connections occurring in adjacent levels.
However, despite being an intuitively appealing concept, the hierarchical organization of large networks,
such as those in the brain, are difficult to identify, especially in absence of additional information
beyond that provided by the connectome. 
In this paper, we propose a framework to uncover the hierarchical structure of a given network, 
that identifies the nodes occupying each level as well as the sequential order of the levels.
It involves optimizing a metric that we use to quantify the extent of 
hierarchy present in a network. 
Applying this measure to various brain networks, ranging from the nervous system of the nematode
\textit{Caenorhabditis elegans} to the human connectome, we unexpectedly find that they exhibit
a common network architectural motif intertwining hierarchy and modularity.
This suggests that brain networks may have evolved to simultaneously exploit the functional advantages of
these two types of organizations, viz., relatively independent modules performing distributed processing
in parallel and a hierarchical structure that allows sequential pooling of these multiple processing streams.
An intriguing possibility is that this property we report may be common to information
processing networks in general.
\end{abstract}

\maketitle


\section{Introduction}

The brain exhibits complexity at multiple levels, 
from individual neurons interacting with neighboring cells, to the emergence of 
large-scale spatiotemporal patterns of activity in entire brain areas, ultimately giving rise to behavior
and cognition~\cite{Lynn2019}.
While the size and scale of nervous systems vary widely across species, it is striking that they
nevertheless perform the key function of enabling the organism to respond appropriately
to an ever changing environment~\cite{HerculanoHouzel2017}. 
The nervous system of the
nematode \textit{Caenorhabditis elegans} comprising $\sim 300$ neurons 
lies at one end of this spectrum, while mammalian brains with tens of billions of neurons
straddle the other extreme. Thus, the complexity of the brain does not
simply arise from its size alone, e.g., the number of constituent neurons,
but is also associated with the connection topology of the wiring between its constituent units~\cite{Bassett2017}. 
Understanding this structural organization, that underpins brain function, requires identifying general design principles that can provide a conceptual scaffolding for describing the connectome.
Here we focus on \textit{modularity} and \textit{hierarchy}, attributes that
have often been associated with structural and/or functional features of brain organization.
However, as these terms have been used with very
different connotations depending on the context [e.g., see Refs.~\cite{Ravasz2003,Chatterjee2007, Bassett2008,Crofts2011,Hilgetag2020} for instances of distinct ways in which hierarchy has been interpreted], their explanatory power has been limited.
While the use of graph theoretic concepts has contributed to a rigorous and widely used
framework for understanding modularity~\cite{Newman2006}, even within the specific arena of
network neuroscience there has been a multiplicity of approaches that seek to 
quantitatively characterize hierarchy.

%
Hierarchical organization has often been inferred, e.g., in the Macaque visual cortex~\cite{Van1983, Lamme1998, Markov2014}, by
observing how information flows across a sequential arrangement of layers,
such that each successive layer integrates the signals obtained from the preceding layer
and performs more complex information processing, thereby defining a bottom-up flow.  
In parallel, top-down feedback connections from higher processing levels to those at lower levels 
implement control mechanisms that allow
adaptation and fine-tuning of responses~\cite{Rao1999,Urgen2015}. Similar hierarchical
organization has been reported in different species~\cite{Coogan1990,DSouza2016} as well as other sensory modalities~\cite{Hackett2014}.  
Indeed, networks in general that are involved in complex information processing 
appear to be characterized by such an arrangement of reciprocal connections between 
nodes belonging to successive levels occurring in a sequence~\cite{Ispolatov2008,Davidson2010,Josephs2022}. 
%
This can, in principle, be related to the optimal use of computational
resources by having successive layers receive information 
appropriately processed 
so as not to overwhelm the handling capacity of the 
constituent nodes. Distributing the processing task across the system, so that
various steps are performed sequentially at successive layers, increases the robustness of the
network against congestion-driven failure arising from bottlenecks that could result from increased computational load at a key node.
In the brain, hierarchical organization 
is also hypothesized to be crucial for coordinating complex sequential behavior~\cite{Abeles1991,Seung2012}, e.g.,
HVC neurons firing in a precise temporal order during the ``song''
of the zebra finch~\cite{Long2010}.

The lack of an universally accepted quantitative measure for hierarchy has meant that almost all 
earlier reports of hierarchical architecture in brain networks have relied on identifying the
distinct layers through additional information about the attributes of the constituent nodes, such
as their function and/or anatomical characteristics. However, in instances where such auxiliary
knowledge is unavailable or incomplete, we require a procedure for 
unambiguously reconstructing the hierarchical sequence of layers from the connection topology alone.   
With rapid progress in methods to determine structural and functional brain networks in
recent times, there is rising interest in
devising novel approaches for analyzing the resulting abundance of connectome data
to identify the inherent organizational features of such networks, in particular, hierarchy~\cite{Harris2019}. As already implied above, 
hierarchical networks share the structural characteristic that their nodes are 
sequentially arranged into several layers with the densest connections occurring between
successive layers. The presence of such an organization which intuitively has functional implications in 
terms of directing the flow along the network, especially for systems involved in processing
information, can be used to fashion a quantitative metric of hierarchy. 
Thus, the hierarchical architecture of a network can in principle be disentangled by identifying the arrangement of all $N$ nodes (say) of a network into 
$L$ levels $l_i$ ($i = 1,\ldots,L$) such that the number of connections between successive levels is maximized
across all possible choices of (a) the number of levels $L$, (b) the sequential arrangement of 
$l_i$ and (c) the level membership of each individual node.

In this paper we propose a general framework based on this insight to identify the 
mesoscopic structure (in particular, the presence of a hierarchical organization) in a network
exclusively from information about its connections and apply it to
multiple connectomes.
We define a metric, the {\it hierarchy index H}, such that 
maximizing it yields the optimal hierarchical decomposition of a network in the sense
as described above.
This is a non-trivial search problem, as even for a fixed value of $L$,
the number of possible ways in which $N$ nodes can be partitioned among the levels is
given by the Stirling number of the second kind, yielding 
an astronomically high number of possibilities. For instance, a connectome comprising
$200$ brain regions can be arranged among $10$ levels in more than
$10^{193}$  different ways.
We solve this combinatorial optimization problem by introducing a heuristic simulated annealing 
routine developed specifically for 
identifying the optimal partitioning of the network components into levels and their corresponding sequential arrangement. 
Benchmarking was carried out on synthetic networks with embedded hierarchical organization
to establish that the hierarchy is correctly identified by the algorithm consistently across realizations.  

Applying this method to various connectomes, ranging from the macro-scale, consisting of tracts linking brain areas, to 
the micro-scale, comprising synapses and gap-junctions between neurons, we identify a robust
meso-scale feature, viz., {\it modular hierarchy}.  
Indeed, our results suggest that the organization of brain networks is characterized by 
an interplay between
the two prominent mesoscopic structural features, viz., modularity and hierarchy, 
such that
neither can independently explain 
the trajectory of signals flowing through the nervous system, relayed from layer to layer and module to module.
Note that, this concept is distinct from that of hierarchical modularity~\cite{Ravasz2002, Guimera2005, Sales2007, Clauset2008,Kunin2022}, which has been used 
in the literature to refer to a nested arrangement of modules.
We show that the layered structure characterizing each module is not completely
independent of that in other modules, suggesting a weak sequential order among the
modules themselves rather than a dominant global hierarchy. 
Such an organization is consistent with the functional requirements of the nervous system which processes information in a segregated manner along specialized streams but eventually
requires an overall integration. 
Taken in conjunction with
recent experimental observations indicating that modules that are established initially
subsequently get concatenated~\cite{Murakami2022}, our results suggest that this structural feature may well be developmentally programmed, pointing again to its potential functional relevance. 
\begin{figure}[t!]
\includegraphics[width=1\linewidth]{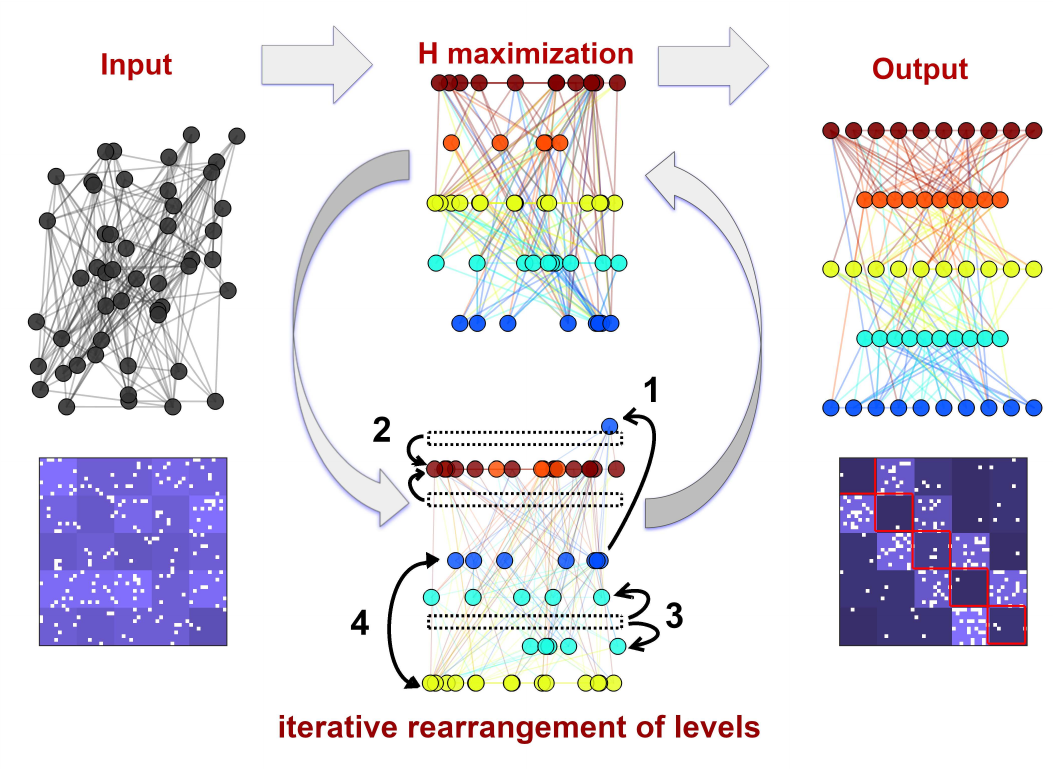}
\centering
\caption{\textbf{Schematic representation of the process for identifying the
underlying hierarchical organization of a connectome.} 
The iterative application of the algorithm proposed here allows 
the innate layered structure of a network which may not be apparent \textit{a priori}
(as indicated by the graph representation and corresponding adjacency matrix
of the input, left) to be made explicit (shown in graph and matrix representations
of the output, right), enabling the identification of the hierarchical organization.
At each iteration, the process incrementally maximizes
the hierarchy index $H$ (middle) by improving an initial assignment of
the nodes into a number of levels by performing any one of four types of random rearrangements
(selected with different probabilities, see Methods) : 
(1) moving 
a node to a different level that is either chosen randomly from the
existing ones or is newly created, (2) merging two randomly chosen levels, 
(3) splitting a randomly chosen level into two adjacent levels,
and (4) exchanging the order of a pair
of randomly chosen levels in the sequence. 
The hierarchical structure of the network in the output is apparent from the dense 
connections in the blocks immediately adjacent to the diagonal blocks, 
representing relatively high connectivity
between nodes occurring in adjacent levels.
}
\label{fig:Fig1}
\end{figure}

\section{Results}
The hierarchy index $H$ that we define here (see Methods)
attains its highest possible value for 
a partitioning
of the nodes into a sequence of levels that maximizes the 
density of links between adjacent levels.
Thus, uncovering the hierarchical structure of a network is framed as a combinatorial
optimization problem
that we solve using simulated annealing
(Figure~\ref{fig:Fig1}, see Methods for details).
To establish the effectiveness of the proposed method in identifying the underlying
hierarchical structure of a network, we first apply it to generated ensembles of benchmark random networks
where such an organization is present by design, and compare the inferred sequence
and composition of levels with that known \textit{a priori}. 
The process by which links are
assigned between different nodes allows us to specify the extent of hierarchical
organization, parameterized
by the ratio $h$ of the densities of connections between consecutive levels to 
that between all other levels (as well as, within each level).
This allows us to smoothly vary the nature of the constructed networks from ones
where hierarchy is completely absent ($h=1$), with the connections being 
uniformly distributed throughout the network, to those that are
rigidly hierarchical ($h=0$), with nodes at any level allowed to connect
only with those in levels immediately above or below them (see Methods for details).
The benchmark networks can be decomposed into an optimal set of partitions 
by maximizing $H$ which allows us to recover the mesoscopic topological organization 
embedded in the network to a remarkable accuracy, as measured
by the normalized mutual information between the original and reconstructed
hierarchical configurations. As expected, the performance of the algorithm declines
as the hierarchical character of the network becomes less pronounced (for $h>0.1$), with a reduction
in the similarity between the partitions identified by the algorithm and those inserted by
construction (see S1 Figure in Supplementary Information).

Having validated the accuracy of the proposed hierarchical decomposition on
benchmark networks, 
we apply
this method to uncover any underlying hierarchical organization that
may exist in several connectomes that vary in size and complexity, as well as, 
the scale of resolution of the network. These include the neuronal network
corresponding to the somatic nervous system of the nematode \textit{Caenorhabditis elegans}~\cite{Cook2019}, 
an aggregated network of brain regions compiled from numerous tractographic studies of the Macaque brain~\cite{Stephan2000,Stephan2001,Kotter2004,Modha2010}, and structural brain networks of multiple human subject obtained
through diffusion tensor imaging~\cite{Nooner2012,Brown2012} (see Methods for details).
We describe below the results of the hierarchical decomposition carried out on 
these networks. We begin with the Macaque macro-connectome, whose modular organization
and resulting functional consequences we have recently investigated in detail~\cite{Pathak2022}.

\begin{figure*}[t!]
\centering
\includegraphics[width=0.8\linewidth]{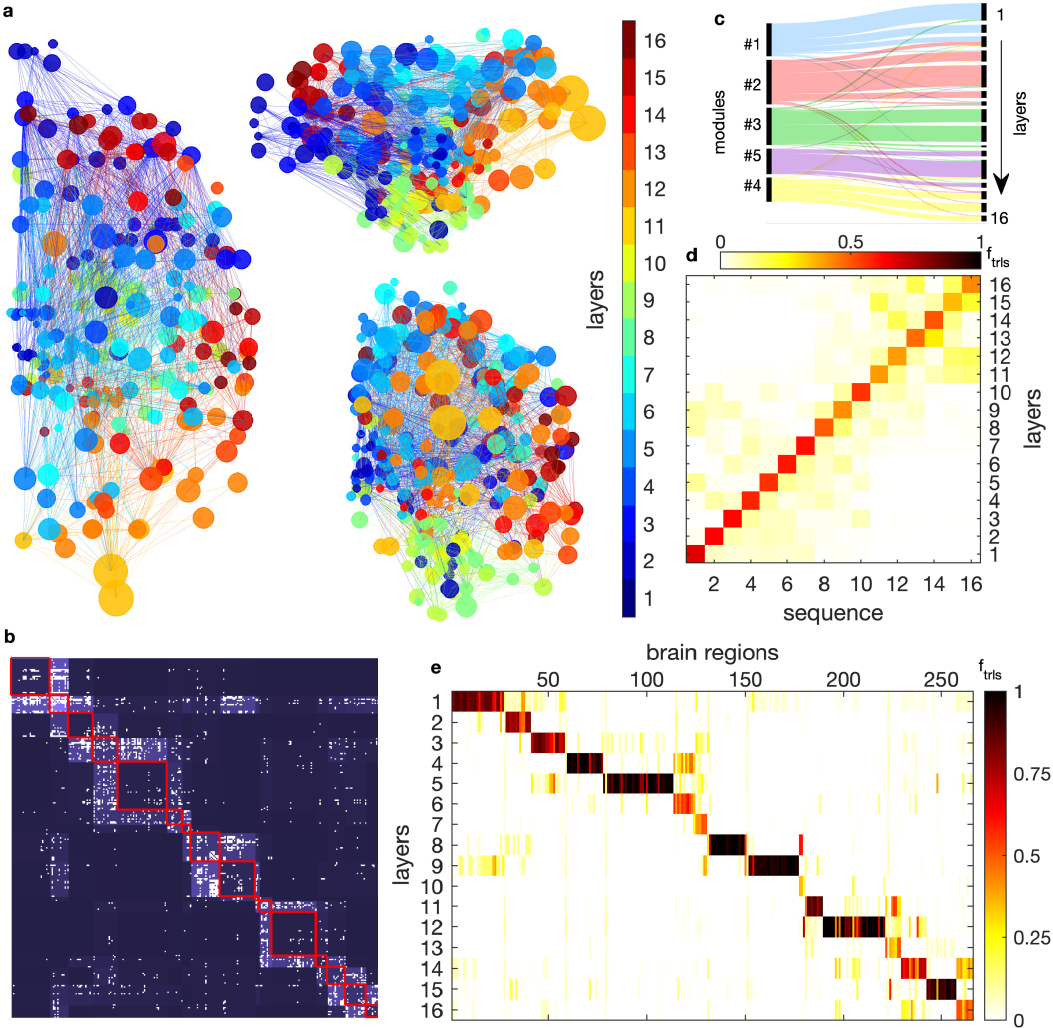}
\caption{
\textbf{The hierarchical organization identified in the macaque connectome.}
(a) Network of brain areas, shown in horizontal (left), sagittal (top right) and coronal 
(bottom right) projections, with the links representing directed axonal tracts between
the areas. The layer in the hierarchy to which an area
(filled circle) belongs is indicated by the corresponding node color (see color key),
while the node size provides a representation of its relative volume.
Each directed link between a pair of areas has the same color as that of the source node.
(b) Adjacency matrix representation of the macaque connectome, with nodes
(brain areas) arranged according to the hierarchical level in which they occur in
the decomposition shown in (a).
The existence of a directed connection between a pair of brain areas $i,j$ is represented 
by the corresponding entry in the matrix being colored white.
The density of connections between areas belonging to the same or different hierarchical levels
is indicated by the brightness of the corresponding block.
(c) Alluvial diagram representation of the association between the modules (left) and the
hierarchical layers (right) to which the different areas belong.
(d) The robust sequential arrangement of the layers is indicated by the 
relative frequency $f_{trls}$ of each layer in a reference sequence
(ordered along the ordinate) occurring 
at specific positions (shown along the abscissae) in the hierarchical decomposition obtained in each of $10^3$ realizations.
The reference sequence is the hierarchical decomposition shown in (a).
%
%
%
(e) The invariance of the hierarchical partitioning of the brain areas identified
across different realizations is quantified by the relative frequency $f_{trls}$ with 
which an area occurs at a given layer ordered as per the reference hierarchical arrangement 
shown in (d). 
}
\label{fig5_4}
\end{figure*}
A typical hierarchical decomposition of the Macaque connectome obtained by
applying our algorithm is shown in Fig.~\ref{fig5_4}~(a-b).
The specific sequence displayed here comprises $16$ levels into which the brain regions 
are arranged and is taken as the reference sequence (see S1 Table in
Supplementary Information) against which we compare 
other decompositions generated by multiple realizations of the algorithm (see Methods).
As the diameter of the network ($=8$) is much smaller than the number of layers 
into which the connectome is seen to be partitioned in these decompositions,   
it may initially appear counter-intuitive that the shortest path length connecting the two
regions furthest in terms of the distance measured along the network is not
comparable to the separation between the terminal layers in the hierarchical chain.
Such a feature suggests a marked
deviation from a strict hierarchy (characterized by links existing exclusively between
neighboring layers) and has always proved challenging to
any effort at describing the brain in terms of a serial arrangement of layers that
successively process information~\cite{Hilgetag2020}.
This apparent incongruity arises because of a profusion of ``short-cuts'' linking regions that 
lie in layers that
are far apart along the sequential arrangement. This can be established by observing how
the diameter of synthetic networks having comparable number of nodes and layers as the 
connectome, decreases as the density of links connecting non-consecutive layers increases
(see S3 Figure in Supplementary Information). Indeed, it is the presence of these
connections which obscures the underlying hierarchical arrangement of brain networks,
a problem that has been overcome by the hierarchical decomposition method introduced here. 

Almost all the decompositions exhibit a spatially contiguous arrangement in that the
sequentially adjacent levels also appear to be spatially adjacent.
We observe that the levels exhibit a cyclic progression from
the anterior to posterior before eventually turning back. As see from 
the sagittal section [top right panel of Fig.~\ref{fig5_4}~(a)], the sequence begins at 
the pre-frontal cortex (nodes in layers 1 to 3) and  then moves
across the parietal lobe (layers 4 to 7) down to the sub-cortical regions (layers
8 to 10) before proceeding up again to the occipital lobe (layers 11 to 13). The 
subsequent levels then progress in the reverse direction [see the horizontal
section in the left panel of Fig.~\ref{fig5_4}~(a)] across the
temporal lobe (layers 14 and 15) to finally terminate in the pre-frontal cortex (layer 16).
Thus, the terminal levels of the hierarchy are both located in the frontal lobe. 
Fig.~\ref{fig5_4}~(b) shows that most of the connections between brain regions
tend to be concentrated between consecutive layers  (whose nodes occur within the partitions indicated by the red bounding lines) in the hierarchical sequence, consistent with the
intuitive notion of hierarchy that we outline earlier. As can be seen, the sizes of the
layers, measured by number of regions that belong to each of them, are highly
variable, ranging from $3$ (Layer 10) to $36$ (Layer 5).
However, as the brain regions themselves occupy very different spatial volumes,
spanning several orders of magnitude, the size differences between the layers in terms of number of
regions may not easily translate to variation in their spatial scale. 

As noted earlier, it has already been shown that the Macaque connectome 
has a prominent modular organization of the brain regions, defined by communities 
characterized by dense intra-connectivity that are spatially localized to a large extent~\cite{Pathak2022}. 
Fig.~\ref{fig5_4} (c) shows how the two mesoscopic organizational features of the brain,
viz., modularity and hierarchy, relate to each other. 
As can be seen, each module comprises brain regions that largely belong to sequentially adjacent
hierarchical layers, such that we can categorize the network as one that is composed of \textit{modular hierarchies}. In other words, the connections can be partitioned into several modules, 
each of which can be further decomposed into a series of hierarchical layers. We note that
the hierarchy is defined not only in terms of the sequence of layers within each module,
but the different modules themselves occur in the decomposition in a specific order.

As the modularity of the network can potentially interfere with the determination of
the hierarchical sequence, given that both types of mesoscopic organization are based on
the differential connection densities within and across partitions, we have carried out
another series of benchmark tests of the decomposition algorithm 
aimed specifically at such networks whose nodes
are arranged into modules, as well as, hierarchical layers. 
For this purpose, we have constructed an ensemble of synthetic random networks
with a hierarchical structure as determined by the parameter $h$ (defined above),
and whose modular character is parameterized by the ratio $r$ of the densitities
of connections between nodes belonging to the same community ($\rho_o$)  and
those belonging to different communities ($\rho_i$)~\cite{Pan2009}.
Benchmark networks are obtained for given pairs of values of $h$ and $r$, comprising
several modules whose nodes are in turn arranged into multiple sequentially
arranged layers, i.e., the networks embody a modular hierarchical architecture (see S2 Figure in Supplementary Information).
The networks are then decomposed
by maximizing $H$ and the partitioning thus obtained can be compared with the embedded
structure by computing the mutual information between them. We observe that the algorithm uncovers the underlying hierarchical organization in the presence of modules, when both the hierarchical and modular characters of the network are prominent
(i.e., for low $h$ and $r$). 
We note that the embedded organization can be detected 
with an accuracy
that is comparable to that obtained for the exclusively hierarchical synthetic network 
ensembles (described above) 

As the process for partitioning of the network into hierarchical layers (and modules)
is stochastic in nature, different realizations of the decomposition can result in
distinct sets of network partitions, and the sequence in which they are arranged may also
vary. A network with an inherent hierarchical organization should display broad consistency 
across the various decompositions, both
in terms of the membership of the different layers as well as their sequential order (see Methods).
Fig.~\ref{fig5_4} (d) shows that there is strong agreement between the different sequences
of layers in the Macaque connectome obtained from multiple realizations, as indicated by
the prominent diagonal [see also S4 Figure (a-c) in Supplementary Information]. 
This implies that specific levels in the reference sequence (ordinate) 
occur with high relative frequency at corresponding positions in the sequences obtained from the other decompositions (abscissa).
This is complemented by Fig.~\ref{fig5_4}~(e), which shows that the relative 
frequency with which a particular brain region occurs in a specific layer across different realizations
is strongly localized to a single partition. 
In other words, the layer memberships of the brain regions are highly correlated, i.e.,
if a pair of brain regions belong to a particular layer in a sequence (considered as the reference), 
then they will co-occur in a layer in other hierarchical decompositions of the network with
very high probability.
Thus, the Macaque connectome shows a very robust hierarchical organization embedded
within the inherent modular structure of the network, with the composition of individual
layers and the order in which they occur sequentially being largely invariant across
realizations of the decomposition algorithm.
\begin{figure}[htbp]
\includegraphics[width=1\linewidth]{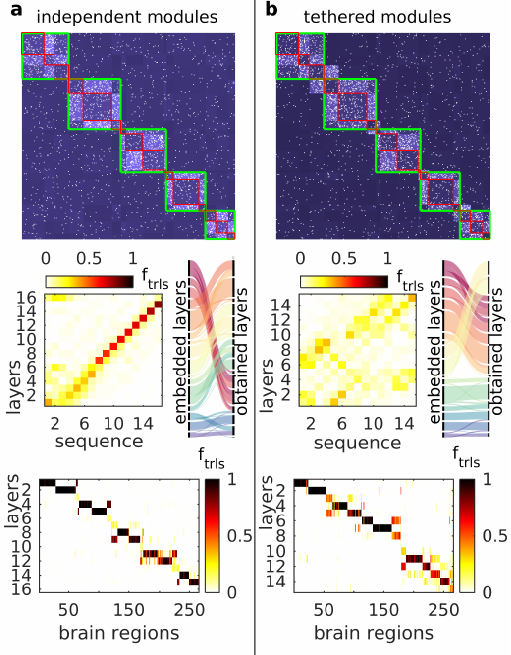}
\caption{\textbf{Inferring the relation between hierarchical layers and modular organization
in the Macaque connectome by comparing the results obtained by partitioning two
classes of synthetic modular networks with embedded layers.}
The benchmark networks correspond to (a) those 
in which each of the modules have a hierarchical organization independent of the
other modules, and (b) those in which the hierarchical levels across
the modules follow a globally ordered sequence, respectively.
In each case, the top panel is an adjacency matrix representation of the network. 
The modules and hierarchical levels
are indicated by green and red bounding lines, respectively, in each matrix.
In the central panels, the extent to which the sequential arrangement of layers is consistent
across $200$ realizations is indicated by the 
relative frequency $f_{trls}$ of each layer in a reference sequence
(ordered along the ordinate) occurring 
at specific positions (shown along the abscissae) in the respective hierarchical decomposition.
The reference sequence for each ensemble is chosen to be the realization that
is most similar (quantified by normalized mutual information) to all other realizations of the hierarchical decompositions of the corresponding benchmark network.
The alluvial diagram representations show
the association between the layers embedded in the benchmark networks (left, considered
to be identical to those in the empirical network) and the layers
obtained by hierarchical decomposition upon application of the proposed method (right).
In the bottom panels, 
the invariance of the hierarchical partitioning of the brain areas identified
across different realizations is quantified by the relative frequency $f_{trls}$ with 
which an area occurs at a given layer ordered as per the corresponding reference hierarchical decomposition.}
\label{fig_benchmark_compare}
\end{figure}

By identifying a network architecture that can reproduce observed features arising from the relation
between modular and hierarchical characteristics of the connectome, we can get vital clues about its organizational principle.
In particular, it will have to demonstrate how a robust, globally sequential ordering of the
identified layers can be consistent with the embedding of these 
layers into prominent modules, which are by definition relatively independent of each other.
With this aim in mind we have considered two classes of synthetic networks possessing both 
modularity and hierarchy. 
One of these classes represent networks that are characterized by 
independent modular hierarchies [Fig.~\ref{fig_benchmark_compare}~(a)].
In these networks, there is a definite sequential ordering of the layers within each module, but not 
of the modules themselves. As the relatively sparse number of connections 
between the different modules may connect nodes at any level in a given module to those occurring
at any level in another, in principle the modules can be placed in an arbitrary order without 
disrupting the hierarchy of the network as a whole.
To contrast with this, we consider another ensemble of networks in which there
is not only a strict ordering of layers within each module, but also across the modules
 [Fig.~\ref{fig_benchmark_compare}~(b)].
In other words, each module is tethered to a specific position in the sequence relative to the other
modules, which arises because most of the connections between consecutive modules in the 
sequence occur between
their respective terminal layers. We note that this implies a rigid sequential arrangement across the modules which would appear to partially contradict the fundamental
attribute of relative independence that characterizes modular structure.
By decomposing these two classes of networks using the algorithm presented here, we
find that while both ensembles are characterized by layer memberships that are consistent 
across realizations, only networks
with independent modular hierarchies exhibit a robust sequential order of the 
identified hierarchical layers, as is observed in the case of the empirical network.
It suggests that the relation between modularity and hierarchy in the connectome
is closer to that represented by networks where the hierarchical arrangement in each
of the modules are relatively independent.
\begin{figure*}[htbp]
\centering
\includegraphics[width=0.8\linewidth]{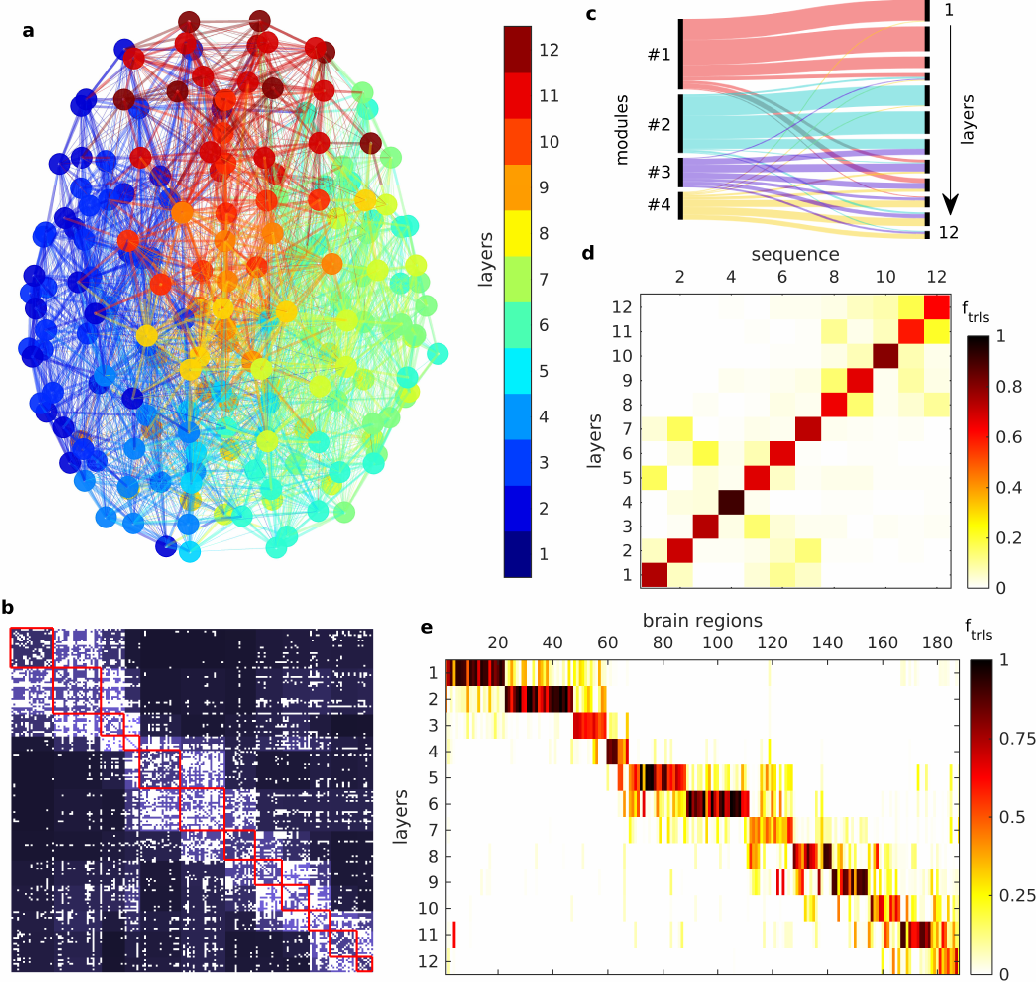}
\caption{
\textbf{The hierarchical organization identified in a human connectome.}
(a) Network of brain areas, shown in horizontal projection, with the undirected links representing axonal tracts between
the areas, for an individual subject in the \textit{NKI/Rockland sample}~\cite{Nooner2012} (see Methods). The layer in the hierarchy to which an area
(filled circle) belongs is indicated by the corresponding node color (see color key).
Each link between a pair of areas is assigned the color of one of the two nodes it joins.
(b) Adjacency matrix representation of the connectome, with nodes
(brain areas) arranged according to the hierarchical level in which they occur in (a).
The existence of a connection between a pair of brain areas $i,j$ is represented 
by the corresponding entry in the matrix being colored white.
The density of connections between areas belonging to the same or different hierarchical levels
is indicated by the brightness of the corresponding block. 
(c) Alluvial diagram representation of the association between the modules (left) and the
hierarchical layers (right) to which the different areas belong.
(d) The robust sequential arrangement of the layers is indicated by the 
relative frequency $f_{trls}$ of each layer in a reference sequence
(ordered along the ordinate) occurring 
at specific positions (shown along the abscissae) in the hierarchical decomposition obtained in each of $200$ realizations.
The reference sequence is the hierarchical decomposition shown in (a).
(e) The invariance of the hierarchical partitioning of the brain areas identified
across different realizations is quantified by the relative frequency $f_{trls}$ with 
which an area occurs at a given layer ordered as per the reference hierarchical arrangement 
shown in (d). 
}
\label{fig5_5}
\end{figure*}
We next investigate the hierarchical organization of a human connectome, obtained
from a representative individual subject
(see Methods). Fig.~\ref{fig5_5}~(a) shows a specific decomposition that is chosen to be
the reference sequence (see S2 Table in Supplementary Information). The
$188$ brain regions across the two hemispheres that comprise the connectome
are seen to be partitioned into $12$ layers.
The regions belonging to consecutive
layers are also physically adjacent as can be observed from their spatial locations
in the horizontal section of the brain shown in Fig.~\ref{fig5_5}~(a).
As in the case of the Macaque, the number of hierarchical layers is larger than the diameter 
of the network ($=4$), which can be attributed to the many connections across non-consecutive
layers functioning as ``short-cuts'' (see S5 Figure in Supplementary Information which shows
the dependence of the diameter of equivalent synthetic networks on the ratio $h$).
Fig.~\ref{fig5_5}~(b) shows that these short-cuts are quite substantial in number. Indeed,
they are more numerous in the human connectome compared to that of the Macaque, which 
can possibly be associated with the much higher overall connection density in the former~\cite{Pathak2020c}.
The network also exhibits modular organization characterized
by the existence of $4$ modules, two of which mostly comprise regions from the left hemisphere
while the other two have a majority of their members in the right hemisphere.
The relation between the compositions of the modules and the hierarchical layers is indicated
by the alluvial diagram in Fig.~\ref{fig5_5}~(c). It suggests that similar to the Macaque
connectome, the network can be viewed as possessing modular hierarchies.
Again as in the Macaque, we see that the network displays a robust sequential ordering
of the layers [Fig.~\ref{fig5_5}~(d), see also S4 Figure (d-f) in Supplementary Information], 
with the identities of the members of each layer
being broadly consistent across different realizations of the hierarchical decomposition 
[Fig.~\ref{fig5_5}~(e)].
We have carried out similar hierarchical decompositions of other human connectomes
obtained from subjects of different ages, which are partitioned by
our algorithm into a similar number of hierarchical layers ranging from $11$ to $14$ 
having robust sequential arrangement, as well as, layer membership 
(see S6 Figure in Supplementary Information).
\begin{figure*}[t!]
\centering
\includegraphics[width=0.8\linewidth]{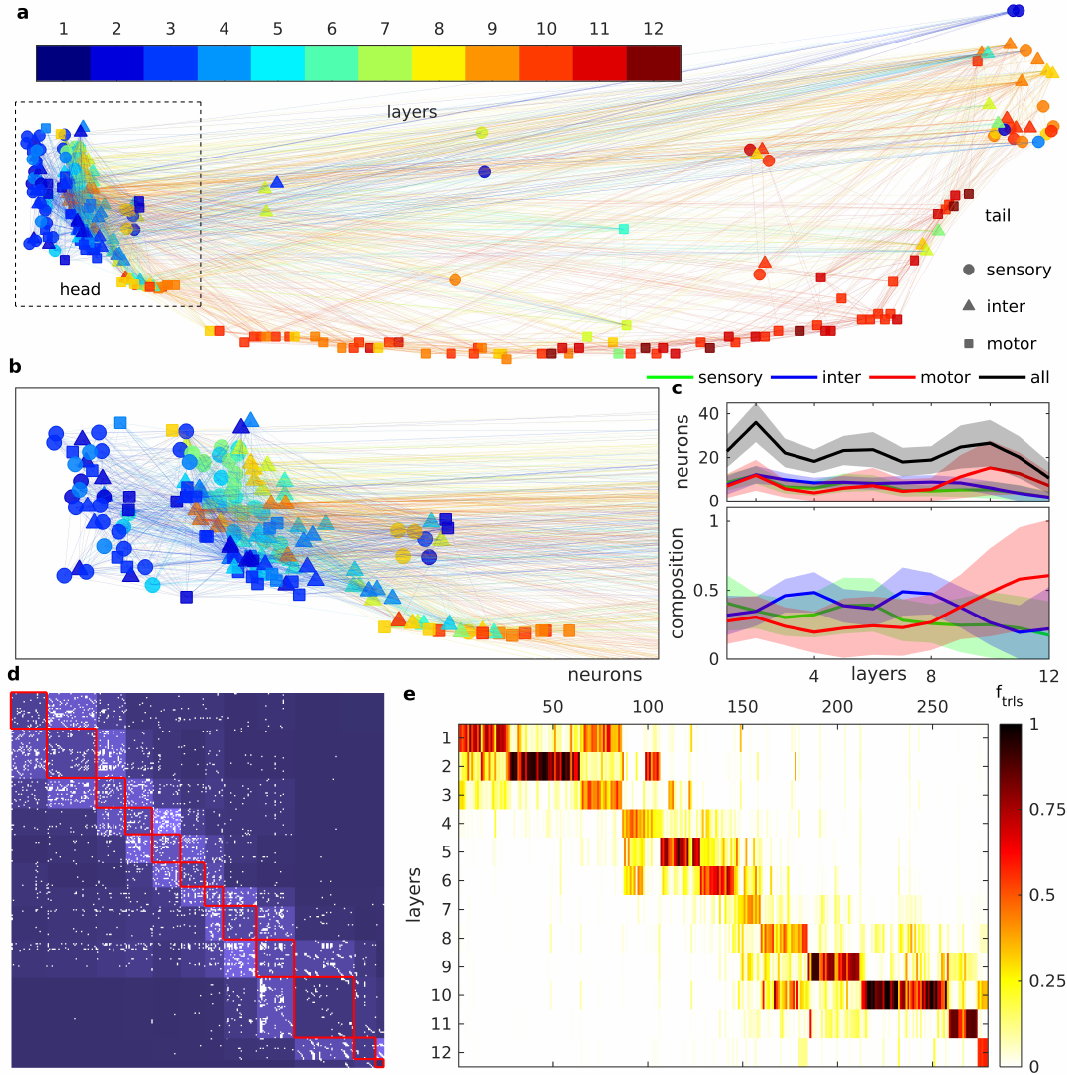}
\caption{
\textbf{The hierarchical structure identified in the somatic nervous system
of the nematode \textit{Caenorhabditis elegans}.}
(a) Spatial representation of the network of synapses between the $279$ connected
neurons that control all activity except pharyngeal movements in the mature
hermaphrodite individuals of the species. The nodes representing the neurons
are arranged according to their position in the worm body along the 
anteroposterior axis, the head and tail being indicated in the figure.
The node color indicates the layer in the hierarchy to which a neuron belongs (see color key),
while the shape indicates whether it is a sensory (circle), motor (square)
or interneuron (triangle).
Each directed synaptic link between a pair of neurons has the same color as the source node.
To resolve the layered organization of the connections between the densely clustered neurons 
in and around the nerve ring near the head, the area enclosed within the broken lines is shown magnified in panel (b). 
(c) The total number of neurons (black), as well as, the individual functional subtypes, viz.,
sensory (green), motor (red) and interneurons (blue) at each level of the hierarchy 
(upper panel), and the fraction of each subtype in these levels (lower panel).
The solid curve represents the mean while the band represents the
dispersion across $200$ realizations of the hierarchical decomposition of the network. 
Note that the sensory neurons are relatively more numerous at the initial layers while 
motor neurons dominate the final layers, with the representation of interneurons
peaking in the middle.
(d) Adjacency matrix representation of the \textit{C. elegans} somatic neuronal network,
with nodes (neurons) arranged according to the hierarchical level in which
they occur in the decomposition shown in (a). The existence of a directed synaptic connection between a pair of neurons $i,j$ is represented 
by the corresponding entry in the matrix being colored white.
The density of connections between neurons belonging to the same or different hierarchical levels
is indicated by the brightness of the corresponding block. 
(e) The invariance of the hierarchical partitioning of the neurons identified
across different realizations is quantified by the relative frequency $f_{trls}$ with 
which a neuron occurs at a given layer ordered as per the reference hierarchical arrangement 
shown in (d). 
}
\label{fig5_3}
\end{figure*}
As a final demonstration of the proposed hierarchical decomposition method, 
we consider the network of chemical synapses connecting $279$ neurons which belong to 
the somatic nervous system of \textit{Caenorhabditis elegans} (see Methods).
A typical partitioning of the network, chosen to be the reference sequence (see S3 Table in Supplementary Information), is shown in Fig.~\ref{fig5_3}~(a),
where the membership of the neurons amongst the $12$ layers that are obtained
for this realization are indicated using different colors. 
As in the case of the networks of brain regions analyzed above, the diameter
of this neuronal network ($=7$) is seen to be lower than the number of layers identified.
This can be imputed to short-cut connections spanning non-adjacent layers  
(see S7 Figure in Supplementary Information).
Considering the spatial positions of the neuronal cell bodies, we
observe that the initial layers are concentrated around the nerve ring located at the head 
of the organism. Subsequent layers have neurons that are located in the tail, while the
neurons of the ventral cord (laid out along the anterior-posterior axis of the worm body
and consisting almost exclusively of motor neurons that coordinate locomotion) 
occupy the final layers in the sequence. Thus, we see a deviation from the spatial contiguity 
of consecutive hierarchical layers that marked the connectomes of macaque and human.
This could possibly be a consequence of many of the neurons having long
processes that span almost the entire length of the organism,
such that their connections are not just confined to the vicinity of the cell body~\cite{celegans2,Pathak2020a}.
Fig.~\ref{fig5_3}~(b) shows the neurons clustered in the various ganglia that are located 
in the head of the organism. In this magnified view, we observe that the 
initial layers broadly appear to be spatially ordered, with their composition being dominated
by sensory and interneurons [indicated by the shape of the symbol representing each neuron,
see key in Fig.~\ref{fig5_3}~(a)]. This is substantiated in Fig.~\ref{fig5_3}~(c), which indicates
the number of neurons of different types that occur in each layer (top panel), as well as
their relative fraction in these layers (bottom panel), averaged over multiple realizations
of the hierarchical decomposition algorithm. The variation in the size of each layer 
(both in terms of the total number of neurons, as well as, of each functional type)
appears to be relatively low across the different partitionings.
The initial layers appear
to have a larger fraction of sensory neurons, while the interneurons predominate the
composition of layers that occur in the middle of the sequence. Motor neurons, on the
other hand, constitute the bulk of the last few layers. 

The adjacency matrix for \textit{C. elegans} neuronal
connectivity shown in Fig.~\ref{fig5_3}~(d) illustrates the dense inter-level 
connectivity between consecutive levels, which is characteristic of a
strongly hierarchical organization. As in the cases of the networks of brain regions described
above, this network also has a modular organization comprising three modules [the association
between their neuronal composition and that of the layers is shown in S8 Figure, left 
panel, in Supplementary Information]. One of the modules comprise neurons
that are mostly located in the ventral cord, while the neurons in the various ganglia are divided amongst the 
other two modules. Each module exhibits a distinct hierarchical arrangement of layers within
them, suggested by the relatively little overlap in the modular memberships of neurons in each
layer. Thus, it appears that the ``modular hierarchy'' organization principle that we observe at 
the scale of connections between brain regions could also be operating at the neuronal scale.  
Also, as in the other networks, the identities of nodes belonging to the different layers are 
consistent across realizations, indicating the robustness of the decomposition in terms of 
layer membership of the neurons [Fig.~\ref{fig5_3}~(e)]. 
While the sequential ordering of the layers shows more variability [see S4 Fig (panels g-i) in Supplementary Information] than that seen in the case of other networks, the reference
sequence is largely conserved across the plurality of the obtained decompositions.
The robustness of the sequence of hierarchical layers is even more pronounced when we augment
the network with additional links (approximately one-third of the number of synapses)
corresponding to electrical gap junctions between neurons [see S9 and S10 Figures in 
Supplementary Information]. Indeed, the network obtained by incorporating these links 
also exhibits hierarchical organization having a similar number of layers,
with the identities of the neurons belonging to each layer remaining consistent across
decompositions. Moreover, the relation between the modules of this network and its 
hierarchical layers also support the hypothesis that the networks comprise 
relatively independent modules, each with an embedded set of hierarchically ordered 
layers [see S8 Figure, right panel, in Supplementary Information].

\section{Discussion}
The mesoscopic organization of a network is expected to reflect its function~\cite{Guimera2007}.
For instance, the necessity of performing multiple independent tasks in parallel, with relatively
low requirement for coordination between them, may favor a modular architecture. 
A network with such a functional requirement can be partitioned into a number of sub-networks, each characterized by high
intra-connection density facilitating recurrent communication between their constituent nodes,
while having correspondingly fewer connections between nodes belonging to different sub-networks.
On the other hand, a hierarchical network may be preferred if the function typically requires 
performing several steps in sequence (such that each step needs to be finished before initiating the next), possibly coordinating across many input streams. 
Such a connection topology would promote efficient serial processing, often in conjunction
with feed-back and feed-forward connection across the levels. 
As we show here, the connection architecture of the brain manifests both of these 
fundamental organizing principles.
Indeed, our analysis of the connectomes suggest a novel structural feature at the
mesoscopic level in these networks that we term modular hierarchies. These are
characterized by the brain regions being segregated into distinct communities, while
at the same time being arranged in a specific sequence of levels within their own community.
The robustness of the modular partitioning, as well as, the hierarchical sequence, suggests
that both of these features are fundamental attributes of the network organization. 
In fact, while there have been previous attempts to identify signatures of hierarchy in the
brain, we venture that it is the simultaneous presence of a strong modular arrangement that
has made such an attempt particularly challenging. The method proposed here is
particularly suited for identifying the interplay of
these two kinds of mesoscopic organization.
We note that a similar architecture is known to be extant in the visual cortex, where 
information coming from different parts of the visual field are processed by different
microcolumns (hence, representing a modular partitioning) with each microcolumn
being composed of a sequential arrangement of neurons that process the information
from the specific part of the visual field received upstream~\cite{Frisby2010}.
Thus, it appears that a common organizing principle may be operating at both the micro-
and macro-scale of the connectome.

It is of interest that the various hierarchical modules themselves appear to be
sequentially arranged, but is consistent with a model postulating that the hierarchical organization
in each of the modules are independent of the others (Fig.~\ref{fig_benchmark_compare}). We note that this parallels recent evidence in mice about the
manner in which the visual network develops, with independent modules being initially established
and subsequently concatenated~\cite{Murakami2022}. It suggests a process by which the
modular hierarchies may be \textit{developmentally} programmed, lending further support to the  
organizational principle suggested by the network \textit{structural} analysis presented here.
In addition, a plausible \textit{functional} relevance of this architecture is hinted at by the
robust sequential relation between the various modules [depicted in S4 Figure in Supplementary Information]. The observations are consistent with the possibility that inter-modular connections
may have a preference for connecting a node in a given layer $a$ of a particular module with
that in layer $b$ of a different module, where these two layers belong to
an overall hierarchical sequence. We note that if such a global arrangement
was strictly enforced, nodes would have been observed to be organized into layers following
a global hierarchical arrangement, which would have been independent of the specific modules to which the individual nodes belong. On the other hand, in the absence of any such preference, the hierarchical level to which a node belongs will have no effect outside of its module and inter-modular
links should be equally likely between any pair of levels. Our results, while not supporting
a global hierarchy independent of the modules, do appear to suggest a certain
preference for sequential arrangement across modules. Functionally, this may provide 
a basis for systematic integration of information and hence allow for distributive processing
in a network that otherwise has a markedly modular organization and hence   
would have appeared to support a segregated (or specialized) mode of processing~\cite{Bullmore2012}.
Indeed, such a relation between modularity and hierarchy has been hypothesized
in the specific case of the visual system, as a possible solution to the ill-posed problem of feature binding~\cite{DiLollo2012}.

Our analysis of the modular hierarchy present in three connectomes, viz., the network of
white matter tracts linking brain areas in the macaque and the human, and that of neurons
in the nematode \textit{Caenorhabditis elegans}, using the novel hierarchy detection algorithm we introduce here,
reveals various facets of the intriguing interplay between hierarchy and modularity.
In the first brain network we investigate, viz., that of a single hemisphere of the macaque, 
our results suggest the existence of two distinct streams 
along which signals could propagate parallel to 
the anteroposterior axis. 
One of these extends from the
frontal lobe to the occipital via the parietal lobe, whereas the other
extends from the occipital lobe to the frontal via the temporal lobe.
As the occipital lobe comprises the primary visual cortex, belonging to one
of the first layers in which sensory information is processed, we  can
distinguish the two pathways identified above tentatively
as ``downstream'' and ``upstream'', respectively. While
the latter may correspond to sensory stimuli being successively analyzed in brain areas
that perform higher level processing, the former can plausibly be involved in 
sending feedback signals back to the initial layers. We note that an analogous
model of bottom-up and top-down processing working in conjunction has been
proposed in the context of vision~\cite{Rao1999,PascualLeone2001,Boehler2008,Urgen2015,Seth2022}.
Indeed, the macaque visual system
has been reported to comprise several stages of hierarchical processing, the
number of layers estimated being
comparable to that obtained using our algorithm (which is distinct from the procedures
used by these earlier studies)~\cite{Felleman1991, Hilgetag2020}.

%
For the individual human connectomes, as in the case of the Macaque, we observe a clearly
identifiable arrangement of brain regions into modular hierarchies, that is consistent across
subjects  in terms of the number of layers and their sequential ordering. Thus, across different
individuals we observe that regions occurring in the terminal layers are located either in 
the frontal or in parietal lobes. However, the progression from
the initial layers can either be from the left to the right hemispheres or the reverse, depending
on the subject whose connectome is being analyzed,
suggesting that the connectivity from the fronto-parietal regions to the left and right
hemispheres is asymmetric.
More generally, the structural organization of the human brain uncovered using the techniques employed here closely resemble the hierarchically layered architectures employed by artificial
neural networks implementing deep learning~\cite{Richards2019}.  These in turn have been inspired by ideas that, in the brain, information is processed through several sequentially arranged
stages of transformation and representation. 
Indeed, recent studies using MEG and fMRI suggest a strong correspondence between object representation in the various layers of designed artificial deep neural networks and 
the hierarchical topography of visual representations in the human brain~\cite{Cichy2016}.
It has long been assumed that the hierarchical organization of the brain is reflected functionally
in a sensory stimulus being represented at different levels of abstraction in the successive layers,
e.g., in the context of the visual system, the various stages respond successively 
to edges, primitive shapes, etc., and eventually to complex forms~\cite{Serre2007}.
The mammalian brain is, from the context of such layered neural network architectures, believed
to implement a deep architecture with many layers, where depth corresponds to the number
of sequentially arranged stages of non-linear operations, applied on the output of the immediately
preceding layers~\cite{Bengio2009}.
Intriguingly, our results show that connectomes exhibit aspects of ``deep'', as well as, 
``shallow'' architecture. Specifically, the various modules that we identify typically comprise $3$-$4$
layers, while the network viewed in totality can be seen to possess a much higher number
of layers, e.g., $15$-$16$ in the macaque, as the modules are themselves arranged in 
a sequential order. Thus, it appears that there may not be a strict dichotomy between
shallow and deep architectures, but rather they are associated with the scale at which
one analyzes the network organization of the brain. 

The connectomes of macaque and human discussed above differ in a fundamental
manner, in that the former is directed while the latter is undirected.
The method proposed here is nevertheless able to detect broadly similar 
hierarchical organizations in both, which points to the robustness of the technique
to the occurrence of directed links. This is important in the context of our analysis of the neuronal network of \textit{Caenorhabditis elegans}, where we focus on the directed network   
comprising chemical synapses. However, the neurons are also connected by
electrical gap junctions, that in principle allow bidirectional communication, and hence
can be viewed as constituents of an undirected network. 
Considering these networks of synapses and
gap junctions together brings up additional challenges as the 
co-occurrence of directed and undirected links in the same network can 
obscure the hierarchical nature of the connections.
Furthermore, the size of the organism is small relative to the typical length scales of the
components of its nervous system, with the longest neuronal processes spanning 
almost the entire body. This implies that there may not be a strict correspondence between 
the sequential arrangement of the hierarchical layers and the spatial
proximity of neurons occurring in neighboring levels, unlike
in the macro-scale networks for macaque and human. 
%

The hierarchical architecture reconstructed from the 
neuronal network (around three-quarters of whose links comprise synapses) 
agrees with our intuitive notion of how signals are processed through the nervous system following
the stimulation of specific sensory organs, encountering in turn, sensory, inter and motorneurons,
the latter serving as actuators for possible muscle activity.
The fact that our algorithm is able to do this despite the specific challenges of analyzing the 
nematode nervous system, not only highlights its effectiveness in determining hierarchical
organization across scales (from macro-connectome of brain areas to micro-connectome of
neurons) but also underscores the ubiquity of the hierarchical architectural plan of the
nervous system.
To conclude, the algorithm that we present here provides a comprehensive method
for uncovering the hierarchical organization of networks appearing in very different
species, effective across scales, nature of links (viz., directed or undirected) and the  
existence of other structural features such as modules.
Indeed, our results provide a new perspective on debates concerning the extent that 
processing in the brain 
is sequential (as in a hierarchically layered system) as opposed to being compartmentalized  
(as in modular systems), 
by suggesting a novel architecture, viz., modular hierarchies, combining 
aspects of both these mesoscopic organization principles.

\section{Methods}
\subsection*{Data}
\noindent
We have considered empirical connectomes comprising unweighted links corresponding
to anatomical tracts between brain regions in the macaque and human, and synapses and gap
junctions between neurons in the nematode \textit{C. elegans}. In cases where the
original data has weights associated with each link, we consider only the adjacency
matrix of the network.
\subsubsection{Macaque}
\noindent
\textit{Connectivity.} We have used a reconstructed macaque structural
connectome comprising $266$ cortical and subcortical brain regions, with $2602$
directed links between them, as
described in Ref.~\cite{Pathak2022}. It is a revised 
version of an earlier database~\cite{Modha2010}, compiled from more that $400$ separate tract
tracing studies catalogued in \textit{CoCoMac}, a comprehensive neuroinformatics
electronic archive~\cite{Stephan2000,Stephan2001,Kotter2004}.

\noindent
\textit{Spatial Information.} The stereotaxic coordinates and the volume of
each brain region in the connectome has been obtained from several sources, including
the website~\url{https://scalablebrainatlas.incf.org/macaque/PHT00}
associated with the Paxinos
Rhesus Monkey Atlas~\cite{Paxinos2000}, as well as, manual curation from the
relevant research literature [for details see Ref.~\cite{Pathak2022}].
\subsubsection{Human}
\noindent
Human brain structural connectomes were chosen from those of subjects in the \textit{Nathan Kline Institute (NKI) / Rockland Sample}~\cite{Nooner2012} repository of diffusion tensor imaging (DTI) data, made publicly available by the \textit{UCLA multimodal connectivity database} at
\url{http://umcd.humanconnectomeproject.org/}~\cite{Brown2012} as undirected
connectivity matrices. 
The $3$-dimensional coordinates locating each brain region in a standardized space
have been obtained from the above-mentioned database.
\subsubsection{Caenorhabditis elegans}
\noindent
\textit{Connectivity.} Information about the directed connections (corresponding to 
synapses), as well as undirected gap-junctions,
between the $279$ connected neurons of the \textit{C. elegans} somatic nervous
system have been obtained from the dataset published in Ref.~\cite{Cook2019}.

\noindent
\textit{Functional type.} Information about the functional type of each neuron, i.e., whether
it is a sensory, inter- or motor neuron, has been obtained from the
database provided in Ref.~\cite{yamamoto91}. 

\noindent
\textit{Spatial information.} Coordinates of each neuronal cell body projected on a
two-dimensional plane defined by the anterior-posterior axis and the
dorsal-ventral axis, were
obtained from the database associated with Ref.~\cite{choe2004}, accessible
online from \url{https://www.dynamic-connectome.org/}. 

\subsection*{Hierarchy index}
In analogy with the intuitive notion of hierarchy as a sequential ordering of items~\cite{Pumain2006},
we consider a network to be hierarchically organized if its nodes can be partitioned among
multiple levels that are arranged in a specific sequence, with a discernible preference for nodes in 
neighboring levels to be connected to each other. 
Thus, specifying the hierarchical organization of a network not only requires
the partitioning of the nodes into different levels
but the sequential order of the levels that maximizes 
the connectivity between adjacent levels must also be identified.
To quantify the extent to which a given network exhibits
hierarchical organization, we introduce a hierarchy index $H$, 
which for a directed, unweighted network whose nodes have been partitioned into a number of 
sequentially arranged levels is defined as:
\begin{equation}
 H = \frac{1}{L}\sum_{i,j}\left[ A_{ij}-\frac{k_i^{in}\cdot k_j^{out}}{L}\right]\cdot(\delta_{l_i,l_j+1}+\delta_{l_i+1,l_j}),
\end{equation}
where $A_{ij}$ represents an element of the adjacency matrix ($=1$ if there is a directed link 
from node $j$ to node $i$, and $0$
otherwise), $L$ ($= \Sigma_{ij} A_{ij}$) is the total number of connections in the network,
$k_i^{in} (= \Sigma_j A_{ij})$ is the in-degree of node $i$, i.e., the total number of
connections received by it  and $k_j^{out}  (= \Sigma_i A_{ij})$
is the out-degree of node $j$, i.e., the total number of its outgoing connections. 
The largest magnitude of $H$ is obtained for a partitioning that maximizes the number of 
connections between adjacent layers in the hierarchical arrangement. 
This is ensured by performing the summation over only those 
pairs of nodes that occur in immediately neighboring layers in a given partition via
the introduction of the
Kronecker delta functions $\delta_{l_i,l_j+1}+\delta_{l_i+1,l_j}$ ($=1$ if levels $l_i$
and $l_j$ to which $i$ and $j$ belong, are adjacent to each other in the
sequential arrangement, and $=0$, otherwise). 
As $k_i^{in} k_j^{out}/L$ is the probability of a connection from $j$ to $i$ in
a homogeneous random network with the same degree sequence as the network
under consideration, the difference with the occurrence frequency in the empirical network (given by the adjacency matrix \textbf{A})
provides a measure of the excess number of links between adjacent layers over that expected
by chance.
Note that it is expected that $H \sim 0$ for a homogeneous, unstructured network.
We would like to point out that the concept of hierarchy that we quantify by $H$
is distinct from that of hierarchical modularity, i.e., network structure characterized by inter-nested
communities~\cite{Ravasz2002, Guimera2005, Sales2007, Clauset2008}.

\subsection*{Maximization of the hierarchy index}
Having defined the hierarchical arrangement of a network to be a partitioning of the nodes into 
$\mathcal{L}$ sequentially arranged levels that maximizes the hierarchy index $H$, we require 
a procedure by which to obtain this arrangement given only the adjacency matrix of a network
comprising $N$ nodes.
Noting that the analogous task of uncovering the community organization of a network
by determining its optimal partitioning that yields the maximum possible value of 
the modularity measure $Q$ 
is known to be NP-hard~\cite{Brandes2007, Clauset2010}, 
we proceed to obtain an approximate solution to the problem of finding the configuration
that maximizes $H$
by using a probabilistic algorithm, specifically, simulated annealing~\cite{Kirkpatrick1983}
[see Fig.~\ref{fig:Fig1} which summarizes the procedure described below].
As the configuration specifies not only the node membership of each level but also the 
sequence in which these levels occur, the heuristic routine for searching the configuration space
needs to explore different partitionings of nodes, as well as, alternate arrangements of levels.
This is achieved by beginning with an initial configuration that comprises an
arbitrary number (typically $5$) of sequentially arranged levels with the nodes
randomly partitioned between them  
and then iteratively altering the configuration by performing any one of the following operations
at each step:
(i) randomly select any one of the $N$ nodes and move it from the level it is occupying 
to any of the other $\mathcal{L}-1$ levels, (ii) create a new level, placed
at the end of the existing sequence, comprising any one of the $N$ nodes extracted at
random from the $\mathcal{L}$ existing levels, (iii) merge two levels that are chosen
at random from the $\mathcal{L}$ levels, (iv) exchange the positions in the sequence
of any two levels chosen at random from the $\mathcal{L}$ levels, and
(v) split any one of the $\mathcal{L}$ levels chosen at random into two, placed adjacent
to each other in the sequence. The total number of possible ways in which these operations can
be carried out is  ($N(\mathcal{L}-1)+N + ^\mathcal{L}C_2 + ^\mathcal{L}C_2 + \mathcal{L} =  $)$N\mathcal{L} + \mathcal{L}^2$ and at each iteration
of the algorithm we choose any one of these with equal probability.

If the hierarchy index computed for the new configuration resulting from the operation 
carried out at a particular step is higher than the $H$ of the existing configuration,
the alteration to the hierarchical structure is accepted. On the other hand, if the change
in the hierarchy index $\Delta H<0$, the new configuration is accepted with a probability
$P \sim e^{-|\Delta H|/T}$, where the parameter $T$ is referred to as \textit{temperature} in
analogy with thermal annealing. In the course of annealing, the temperature is gradually
reduced according to a cooling schedule, viz., 
$T_n = T_0 e^{-\lambda n}$ where $n$ is the number of iterations and $T_0$
is the initial temperature. 
For the results shown in our paper we have chosen
$T_0 = 10$, $\lambda=2\times10^{-6}$ and have carried out the annealing 
for $n_{max}=2\times10^7$ iterations which was sufficient for convergence
of the process for networks having $N \sim 300$ nodes (corresponding approximately to the
sizes of the empirical networks we have considered).
As the temperature decreases, the system tends to spend longer times in a particular
configuration until a new configuration is accepted and we terminate the algorithm
if the configuration has not altered in the preceding $n_{cutoff}$ iterations [we
have set $n_{cutoff}$ equal to $5$ times the total number of possible operations, 
viz., $5$ ($N\mathcal{L} + \mathcal{L}^2)$].

The solutions resulting from applying the algorithm on a given network need not be
unique, as there could be multiple optimal hierarchical configurations
characterized by high values of $H$. Given this degeneracy, multiple
realizations of the hierarchy index maximization process have been carried out to
construct an ensemble of optimal partitionings for each network (e.g., $10^3$ realizations for the Macaque connectome and
$200$ for the human and \textit{C. elegans} connectomes).
A network is determined to possess a robust hierarchical organization if the
solutions comprising the ensemble are mutually consistent in terms of both the
node membership of the different levels, as well as, the sequence in which these levels are
arranged. The process by which the level of agreement between the different optimal partitionings
of a network is quantified is described below (see section on determining the robustness of the partitions and their sequence).

\subsection*{Benchmarking the performance of the algorithm}
In order to show that the maximization of hierarchy index using the procedure outlined
above does indeed uncover the inherent hierarchical structure of a network (if any), we
test the algorithm on ensembles of random networks whose connection topology
has a hierarchical organization by design. To construct a network having a desired extent
of hierarchy, we use the ratio $h = \rho_{nc}/\rho_{con}$ as a tuning parameter,
where $\rho_{con}$ and $\rho_{nc}$ represent the 
density of connections between nodes occurring in \textit{consecutive levels} in the
the hierarchical sequence  and that between nodes occurring in \textit{all other levels} (including
the same level), respectively. The parameter $h$ can vary over the interval $[0,1]$,
with $h=1$ corresponding to a homogeneous network without any hierarchy, while 
for $h = 0$, nodes at each level connect only to those at the levels immediately above
or below, corresponding to a rigidly hierarchical organization. 
Apart from the ensemble of hierarchical random networks, we have also considered an additional
ensemble of modular hierarchical random networks wherein the network comprises multiple
modules or communities, each containing an embedded hierarchical structure.
 
For the ensemble of hierarchical random networks (see Fig.~S1 in SI), we generate
a benchmark network for a given value of $h$, having $N$ nodes equally distributed among 
$\mathcal{L}$ levels by linking
nodes occurring at different levels with the connection probabilities
\begin{equation}
 \rho_{con} = \frac{\rho_{nc}}{h} = \frac{k\cdot \mathcal{L}}{2(1-h)(1-\mathcal{L})(N/\mathcal{L}) + h\mathcal{L}N},
\end{equation}
where $k$ is the average degree of the network. 
For the simulation results reported here, we chose $\mathcal{L} = 4$, with $N = 272$ and $k =10$ 
(similar to the corresponding values for the empirical networks).


To generate hierarchical modular random networks (see Fig.~S2 in SI), we assume that the
$N (=272)$ nodes of the network are clustered into $m (=4)$ modules. Within each module $\mathcal{L}_m (=4)$ hierarchical layers are embedded, such that the network has ${\mathcal L} = m\, \mathcal{L}_m$ hierarchical layers in total. 
Apart from the tuning parameter $h$ for the hierarchy, we also use an additional
parameter $r$ that specifies the extent of modularity or community organization in the network. 
It is defined as the ratio of the density of connections between nodes belonging
to different modules ($\rho_o$) to those occurring in the same module ($\rho_i$)~\cite{Pan2009}. 
The nodes of a network are equally distributed between the modules and are linked
according to the connection probabilities
\begin{equation}
 \rho_i = \frac{\rho_{o}}{r} = \frac{k}{(N/m)(1-r) + Nr},
\end{equation}
where the average degree $k=10$. 
Having determined the mean connection density $\rho_i$ within a module, we can obtain the connection 
probability between nodes occurring in the different hierarchical levels that are embedded within each module
as
\begin{equation}
 \rho_{con} = \frac{\rho_{nc}}{h} = \frac{\mathcal{L}_m (N/m)\rho_i}{2(1-h)(\mathcal{L}_m-1)(N/\mathcal{L}) + h\mathcal{L}_m (N/m)}.
\end{equation}

For each class of benchmark networks (hierarchical and modular hierarchical) 
we generate $20$ adjacency matrices for each value of $h$ 
which logarithmically spans the interval $[10^{-2},1]$.
The algorithm for maximizing the hierarchy index is applied after randomly permuting the order of the
nodes in the adjacency matrix so that its hierarchical structure is no longer apparent. 
The optimal hierarchical configurations of the network that are obtained from multiple realizations
of the $H$ maximization process can then be compared with the original partitioning of the nodes
into levels that have been embedded by design. 
We can quantify how close two different hierarchical decompositions $A$ and $B$ (comprising
$\mathcal{L}_A$ and $\mathcal{L}_B$ number of levels, respectively) of a network are to each
other by computing a similarity score between 
the sequence of levels $\{l^A_i\}^{\mathcal{L}_A}_{i=1}$ and 
$\{l^B_j\}^{\mathcal{L}_B}_{j=1}$ describing the two decompositions.
For this purpose we use the normalized mutual information~\cite{Mackay2003}, viz.,   
\begin{equation}
I_{\rm norm}~(A,B) = \frac{2 \sum_{i} \sum_{j} P({l^{A}_i},l^{B}_j) \ln
[P ({l^{A}_i},l^{B}_j)/ P({l^{A}_i}) P(l^{B}_j)]}{-\sum_{i} P({l^{A}_i})
\ln P({l^{A}_i}) - \sum_{j} P(l^{B}_j) \ln P(l^{B}_j)},
\label{inorm}
\end{equation}
where $P({l^{X}_i})$ is the probability that a randomly chosen node lies in
level $l_i$ in partition $X\in \{A,B\}$,
while $P({l^{A}_i},l^{B}_j)$ is the joint probability that a randomly chosen node
belongs to level $l^A_i$ in partition $A$ but occurs in level $l^B_j$ in
partition $B$ ($i=1, \ldots, \mathcal{L}_A$, and $j=1, \ldots, \mathcal{L}_B$). 

\subsection*{Establishing robust hierarchical structure in the empirical networks}
As mentioned above, the different hierarchical configurations of a network obtained from multiple
realizations of the $H$ maximization algorithm should be similar both in terms of
the node composition of their levels, as well as, the sequence in which the levels occur, for any
hierarchical organization identified in the network to be robust.
The distribution of normalized mutual information $I_{norm}$ (see Eq.~\ref{inorm}) 
between every pair of hierarchical configurations obtained
can give us a gross measure for the variability between the
solutions obtained from the different realizations. However, in order to quantify the extent to which
each node in the empirical network occupies a consistent position in the hierarchical sequence
of levels we need to first identify a \textit{reference sequence} $\mathcal{R}$ with which to compare all
configurations. To this end, for each hierarchical
decomposition $A$, we compute the mean of the normalized mutual information between it and the decompositions $X$
obtained from all other realizations, viz., $\overline{I}_{norm}(A)=\langle I_{norm}(A,X)\rangle_X$,
and choose the configuration that has the maximum value of $\overline{I}_{norm}$ as the reference.

\noindent
\textit{Robustness of nodal composition of the levels.} Upon numbering the levels of the reference decomposition as they occur in sequence from $1,\ldots,\mathcal{L}_\mathcal{R}$,
a mapping $\mathcal{F}_{l}: \{l^X_i\}^{\mathcal{L}_X}_{i=1} \rightarrow 
\{l^\mathcal{R}_j\}^{\mathcal{L}_\mathcal{R}}_{j=1}$ is established between the levels 
in any hierarchical configuration $X$ obtained from the different realizations and those occurring 
in $\mathcal{R}$. 
This is achieved by identifying for each level $i \in X$
the corresponding level in $\mathcal{R}$ with which it has maximum overlap.
Thus, it allows us to express the identity of the level that a particular node belongs to in a given realization in
terms of a standard numbering convention common across all realizations, viz., that of the levels of the reference configuration $\mathcal{R}$.
We then compute the fraction of realizations (or trials) 
$f_{trls} (q,p)$ in which node $p$ occurs in level $q$ of the reference sequence 
[see Fig.~\ref{fig5_4}~(e), Fig.~\ref{fig5_5}~(e) and Fig.~\ref{fig5_3}~(e)].
For a hierarchical organization identified by the algorithm to be considered robust, 
the nodes should occur consistently in the same hierarchical level, implying that
the distribution of $f_{trls} (q,p)$ is highly localized. We have ensured that
for each of the empirical networks that we have investigated here, the
majority of nodes $p \in \{1,2, \ldots N\}$ satisfy ${\max}_q f_{trls}(q,p) > 0.5$.
 

\noindent
\textit{Robustness of the sequence of hierarchical levels.}
While the composition of each of the network partitions that correspond to the different levels may be consistent
across multiple realizations of the hierarchical decompositions, it is possible that the order $\{1, \ldots, \mathcal{L}_X\}$ in which the 
levels occur sequentially in a given realization $X$ may be drastically different
from that of the reference sequence $\mathcal{R}$. Therefore, we need to ensure that the sequential
arrangement of the partitions are also consistent between the different realizations. For this purpose,
we construct another mapping $\mathcal{F}_{s}: \{l^\mathcal{R}_i\}^{\mathcal{L}_\mathcal{R}}_{i=1} \rightarrow \{l^X_j\}^{\mathcal{L}_X}_{j=1}$ which relates the levels in the reference
sequence $\mathcal{R}$ to those occurring in  
the configuration $X$. In contrast to the mapping $\mathcal{F}_{l}$,
this is achieved by identifying for each level $i \in \mathcal{R}$
the corresponding level in $X$ with which it has maximum overlap.
Subsequently, we re-order the $\mathcal{L}_\mathcal{R}$ layers of $R$ according to the rank of 
the layers in $X$ that they map to.
This allows us
to identify the extent to which the sequential arrangement of levels in $\mathcal{R}$
gets rearranged in a realization $X$.
%
%
The robustness of the sequential order of the hierarchical levels is quantified by computing the fraction of realizations (or trials) $f_{trls}(i,j)$  
in which layer $i$ of the reference sequence $\mathcal{R}$ 
occurs in position $j$ of the sequence obtained upon re-ordering according to the mapping with $X$ 
[see Fig.~\ref{fig5_4}~(c) and Fig.~\ref{fig5_5}~(c), as well as, Fig.~S4 and Fig.~S10 in SI].
In the ideal situation, where the hierarchical levels consistently occur in exactly the same sequence 
across all realizations, $f_{trls}(i,j) = \delta_{ij}$, i.e., the Kronecker delta function for $i,j=1, \ldots, 
\mathcal{L}_\mathcal{R}$. Note that for the empirical networks
investigated here, the matrices representing  $f_{trls}$  have most diagonal entries close to $1$ with
off-diagonal entries $\ll1$ indicating that the sequential arrangement of the levels is robust. \\

\begin{acknowledgments}
SNM has been supported by the IMSc Complex Systems Project (12th 
Plan), and the Center of Excellence in Complex Systems and Data 
Science, both funded by the Department of Atomic Energy, Government of 
India. The simulations and computations required for this work were 
supported by High Performance Computing facility (Nandadevi and 
Satpura) of The Institute of Mathematical Sciences, which is partially 
funded by DST.
\end{acknowledgments}

%

\clearpage

\onecolumngrid

\setcounter{figure}{0}
\renewcommand\thefigure{S\arabic{figure}}  
\renewcommand\thetable{S\arabic{table}}

\vspace{1cm}

\begin{center}
{\large SUPPLEMENTARY INFORMATION FOR}

\vspace{0.3cm}
{\Large
A hierarchy index for networks in the brain reveals a complex entangled organizational structure}

\vspace{0.4cm}
Anand Pathak$^{1,2}$, Shakti N. Menon$^{1}$ and Sitabhra Sinha$^{1,2}$

\vspace{0.2cm}
\small{$^1$ The Institute of Mathematical Sciences, CIT Campus, Taramani, Chennai 600113, India \\
$^2$ Homi Bhabha National Institute, Anushaktinagar, Mumbai 400 094, India}
\end{center}

\vspace{1cm}

\noindent {\bf S1 Table. The hierarchical decomposition of the macaque connectome.} 
The different columns of the table display the different brain areas belonging to each of the
hierarchical layers, corresponding to those shown in  Fig.~2~(a) of the main text, identified by the method presented in the main text. The names of the areas are as per the convention followed
in the database of $266$ macaque brain areas used in our study (for details see the 
subsection \textit{Data} in the Materials and Methods section of the main text).

\vspace{.2in}

\noindent {\bf S2 Table. The hierarchical decomposition of a human connectome.}
The different columns in the \textit{Layers} Sheet of the table display the different brain areas belonging to each of the
hierarchical layers, corresponding to those shown in  Fig.~4~(a) of the main text, identified by the method presented in the main text. 
The brain areas indicated by the node numbers are identified in the \textit{Nodes} Sheet of the table
using the convention followed
in the database of $188$ brain areas of a subject chosen from the NKI-Rockland sample 
(for details see the 
subsection \textit{Data} in the Materials and Methods section of the main text).
\vspace{.2in}

\noindent {\bf S3 Table. The hierarchical decomposition of the neuronal network comprising
only synapses of the nematode {\em Caenorhabditis elegans}.} 
The different columns of the table display the different neurons belonging to each of the
hierarchical layers, corresponding to those shown in  Fig.~5~(a) of the main text, identified by the method presented in the main text. 
The names of the neurons are as per the convention followed
in the database of $279$ neurons belonging to the largest connected component of the
somatic nervous system used in our study (for details see the 
subsection \textit{Data} in the Materials and Methods section of the main text).

\vspace{.2in}

\begin{figure*}[t!]
\centering
\includegraphics[width=0.8\linewidth]{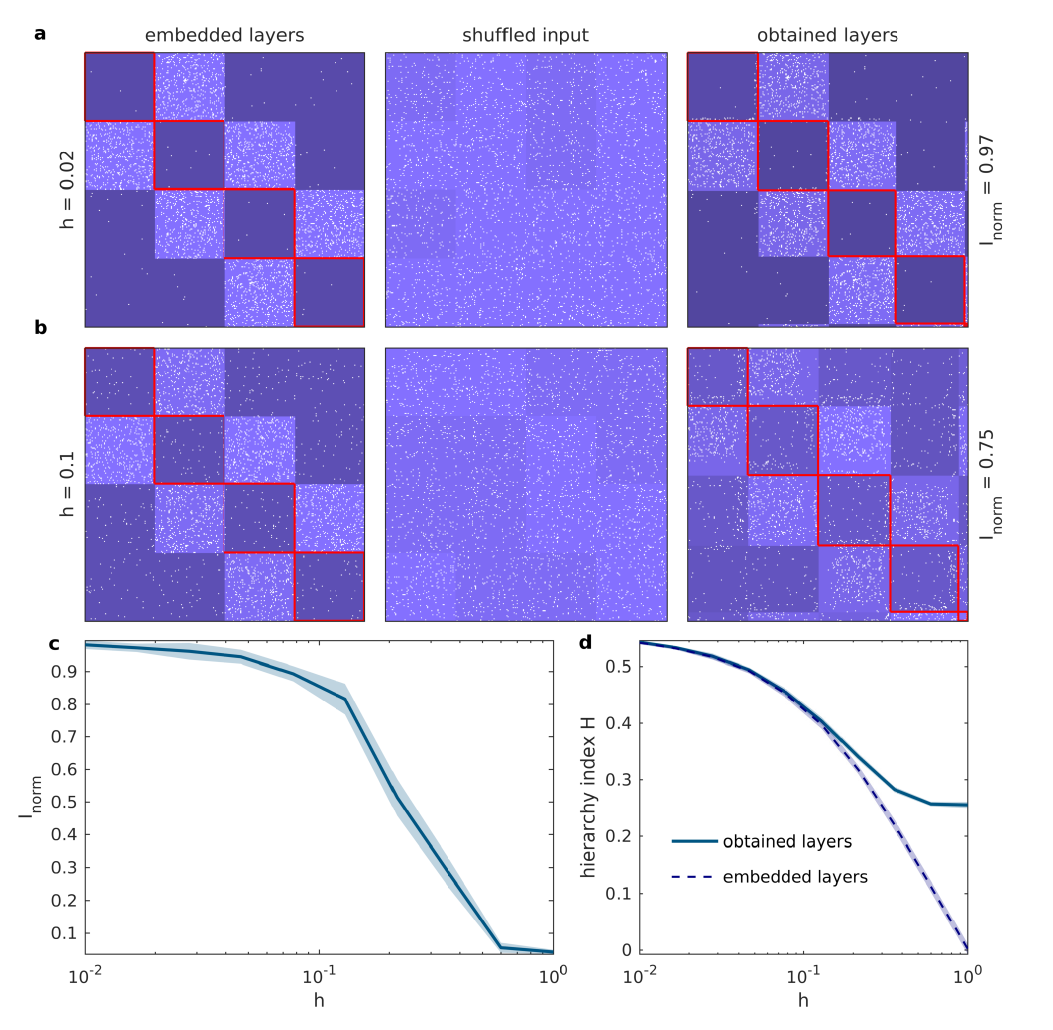}
\caption{
\textbf{Performance of the hierarchy determination algorithm on benchmark networks.
}
(a-b) (Left) Adjacency matrices representing directed random networks having embedded hierarchy
are shown with the different levels (specified by construction, each level being
indicated by red bounding lines) made visually explicit. 
(Centre) The proposed method for determining hierarchical organization is applied
on the matrix after randomly reordering the nodes of the network keeping the connections
invariant so as to obscure the mesoscopic structure.    
(Right) The resulting matrix obtained by rearranging nodes into levels (indicated by red bounding lines) according to the 
proposed method.
The existence of a directed connection between a pair of nodes $i,j$ is represented 
by the corresponding entry in the matrix being colored white.
The density of connections between nodes belonging to the same or different hierarchical levels
is indicated by the brightness of the corresponding block.
The hierarchy of a benchmark network is parameterized by
the ratio $h$ of the connection densities of non-consecutive
levels ($\rho_{nc}$) to that of
consecutive levels ($\rho_{con}$). The results of applying the method on
(a) a strongly hierarchical network ($h=0.02$)  is contrasted with that
obtained from (b) a moderately hierarchical network ($h =0.1$). 
The accuracy of the hierarchical organization determined by the proposed method is quantified by 
computing the similarity, in terms of the normalized
mutual information $I_{norm}$ between the 
partitions separating the embedded levels (known from construction) of the benchmark networks
and those obtained by applying the hierarchy determination method
(see Materials and Methods section in the main text). 
(c) Performance of the hierarchy detection method on benchmark networks ranging
from strongly hierarchical ($h=0.01$) to homogeneous ($h=1$).
Note  that up to $h \sim 0.1$, the hierarchical levels determined by the proposed
method closely match the original embedded levels.
(d) Hierarchy indices $H$ corresponding to the partitioning into
levels obtained from the proposed method (solid curve) compared with that of the original partitions
embedded by construction (broken curve). As expected from (c), the two curves 
almost coincide up to $h \sim 0.1$.
The solid curves in (c) and (d) are obtained by averaging over $20$ network realizations for each value of $h$, the standard deviations being shown as shaded
regions around the curves. }
\label{fig_S1}
\end{figure*}

\begin{figure*}[t!]
\centering
\includegraphics[width=0.8\linewidth]{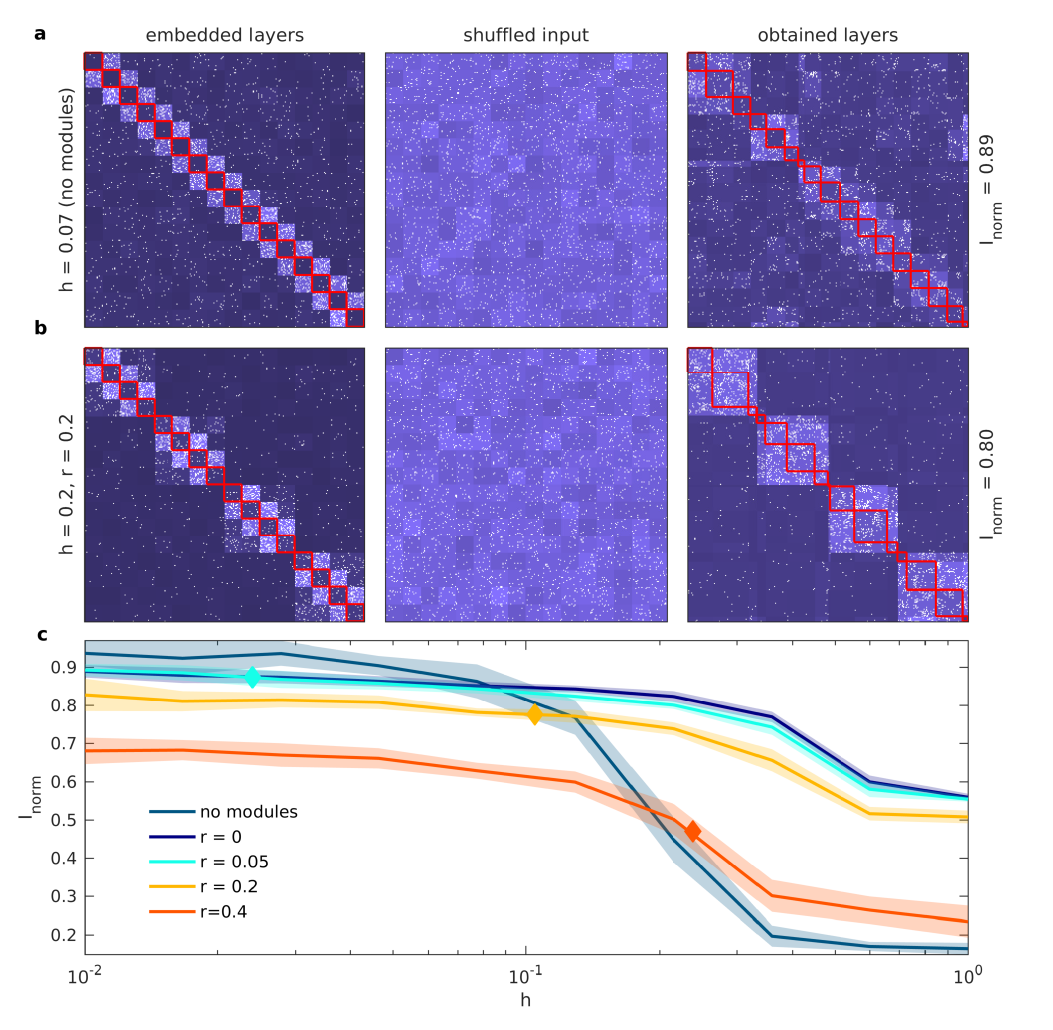}
\caption{
\textbf{Performance of the hierarchy determination algorithm on benchmark networks
having both hierarchical and modular organization.}
(a-b) As in Fig.~\ref{fig_S1}, the panels represent adjacency matrices with (left) the embedded
hierarchical levels shown explicitly (indicated by red bounding lines), (centre) the nodes
randomly reordered to obscure the hierarchical organization prior to applying the proposed
algorithm, and (right) the levels obtained by applying the method to partition the network
(indicated by red bounding lines). 
The existence of a directed connection between a pair of nodes $i,j$ is represented 
by the corresponding entry in the matrix being colored white.
The density of connections between nodes belonging to the same or different hierarchical levels
is indicated by the brightness of the corresponding block.
The hierarchy of a benchmark network is parameterized by
the ratio $h$ of the connection densities of non-consecutive
levels ($\rho_{nc}$) to that of
consecutive levels ($\rho_{con}$), while the modularity is parameterized by
$r = \rho_o/\rho_i$, i.e., the ratio of inter-modular to intra-modular connection densities. 
The results of applying the method on
(a) a strongly hierarchical network ($h=0.07$) with no modular organization is contrasted with that
obtained from (b) a moderately hierarchical network ($h =0.1$) having relatively prominent
modules ($r=0.2$). 
As in Fig.~\ref{fig_S1}, the performance is quantified by 
the normalized mutual information $I_{norm}$ between the 
network partitions known from construction and those obtained by
applying the hierarchy determination method. 
(c) Performance of the hierarchy detection method on benchmark networks having
very different modular characters, ranging from networks comprising several isolated
modules ($r=0$) to weakly modular ($r=0.4$), as well as those without any modules,
shown for a range of the parameter $h$ varying
from strongly hierarchical ($h=0.01$) to homogeneous ($h=1$).
The displayed results are for networks with $N = 272$ nodes, having $16$ levels, $4$ modules and
average degree $\langle k \rangle =10$. Each curve is obtained by averaging over
$20$ network realizations for each value of $h$, the  standard deviations being shown as shaded
regions around the curves.
The diamonds shown on the curves corresponding
to modular networks with $r>0$ indicate the values of $h$ below which $\rho_{nc} < \rho_o$,
such that the intra-modular connections become sparser than the inter-modular connections.
}
\label{fig_S2}
\end{figure*}

\begin{figure*}[t!]
\centering
\includegraphics[width=0.8\linewidth]{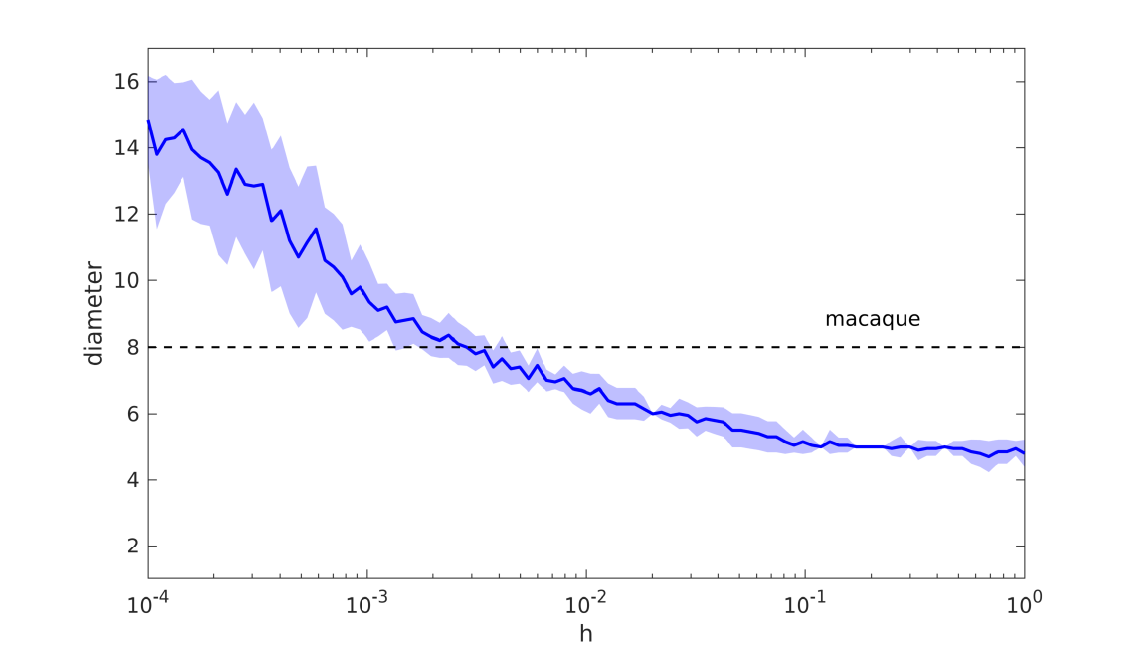}
\caption{
\textbf{Dependence of the graph diameter on the hierarchical
nature of networks with macroscopic attributes identical to the macaque connectome.}
To show that the path length between any two brain areas can be much shorter than the
number of hierarchical levels into which the network is decomposed, the 
diameter, viz., the longest of all the geodesics between pairs of nodes in a network,
is shown for synthetic networks having the same number of levels (viz., $16$) and
number of nodes occupying each level as the macaque connectome (as shown in
Fig.~2 in the main text, see Table S1 for details). 
The diameter is seen to increase as the hierarchical nature of the network becomes more
prominent upon decreasing $h$, viz.,
the ratio of connection densities of non-consecutive
levels ($\rho_{nc}$) to that of consecutive levels ($\rho_{con}$).
The curve is obtained by averaging over
$20$ network realizations for each value of $h$, the  standard deviation being shown as a shaded
region around the curve. For reference, the diameter computed for the empirical
network is indicated by the broken line. 
}
\label{fig_S3}
\end{figure*}

\begin{figure*}[t!]
\centering
\includegraphics[width=0.8\linewidth]{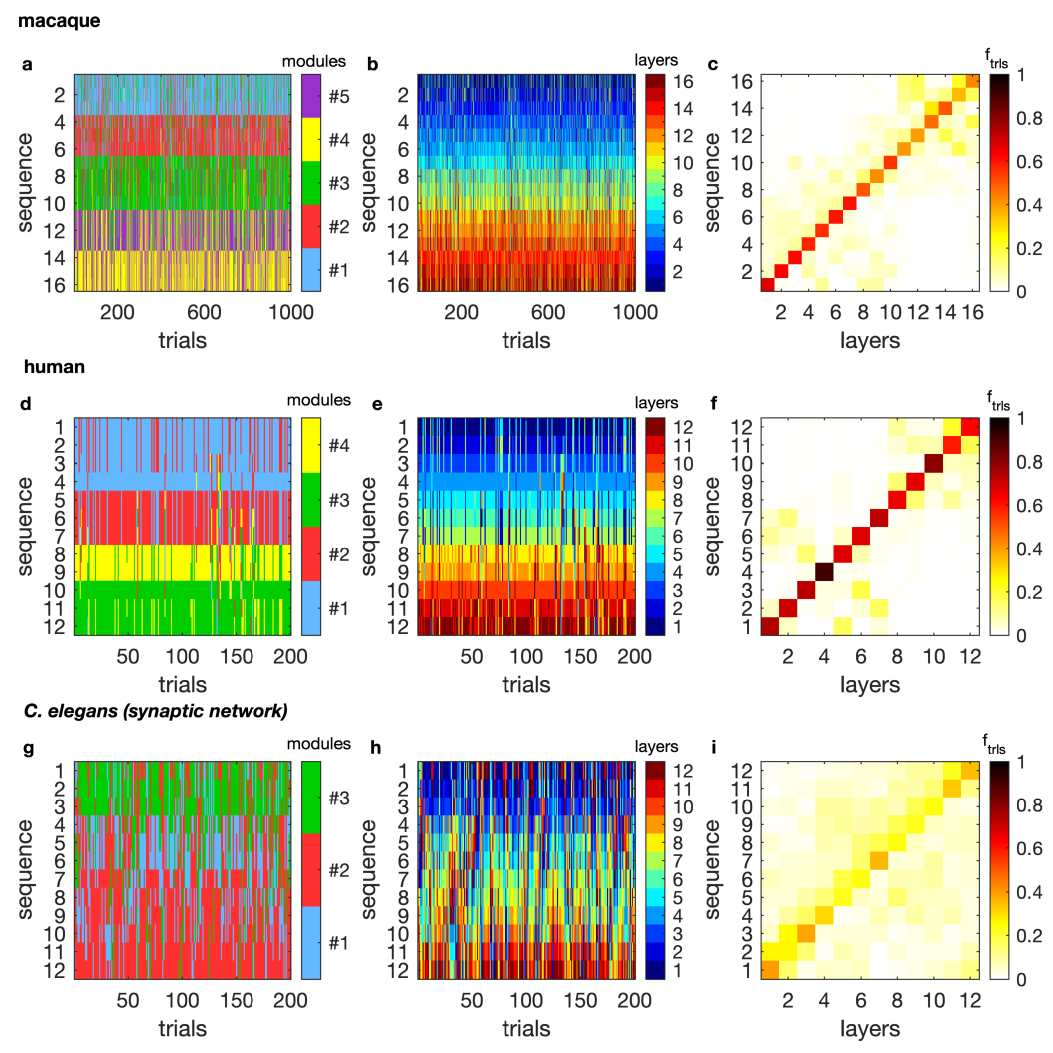}
\caption{\textbf{Robust sequential arrangement of the modules and hierarchical layers in the
macaque, human and nematode connectomes.}
(a-b) The sequence of layers into which the macaque connectome is decomposed in each of $10^3$ 
realizations colored according to (a) the module to which the majority of nodes in each layer belongs, and (b) the position they occupy in a reference sequence, viz.,
the decomposition shown in Fig.~2~(a) in the main text.
(c) The relative frequency of occurrence $f_{trls}$, calculated over $10^3$ realizations, of each layer in the reference sequence (ordered 
along the abscissae) at 
specific positions in the hierarchical decomposition obtained in each trial (shown along the ordinate).
(d-e) The sequence of layers into which the human connectome (analyzed in Fig.~4 in the main text) is decomposed in each of $200$ 
realizations colored according to (d) the module to which the majority of nodes in each layer belongs, and (e) the position they occupy in a reference sequence, viz., 
the decomposition shown in Fig.~4~(a) in the main text.
(f) The relative frequency of occurrence $f_{trls}$, calculated over $200$ realizations, of each layer in the reference sequence (ordered 
along the abscissae) at 
specific positions in the hierarchical decomposition obtained in each trial (shown along the ordinate).
(g-h) The sequence of layers into which the synaptic connectome of \textit{Caenorhabditis elegans}
is decomposed in each of $200$ 
realizations colored according to (g) the module to which the majority of nodes in each layer belongs, and (h) the position they occupy in a reference sequence, viz., 
the decomposition shown in Fig.~5~(a) in the main text.
(i) The relative frequency of occurrence $f_{trls}$, calculated over $200$ realizations, of each layer in the reference sequence (ordered 
along the abscissae) at 
specific positions in the hierarchical decomposition obtained in each trial (shown along the ordinate).
}
\label{fig_S4}
\end{figure*}

\begin{figure*}[t!]
\centering
\includegraphics[width=0.8\linewidth]{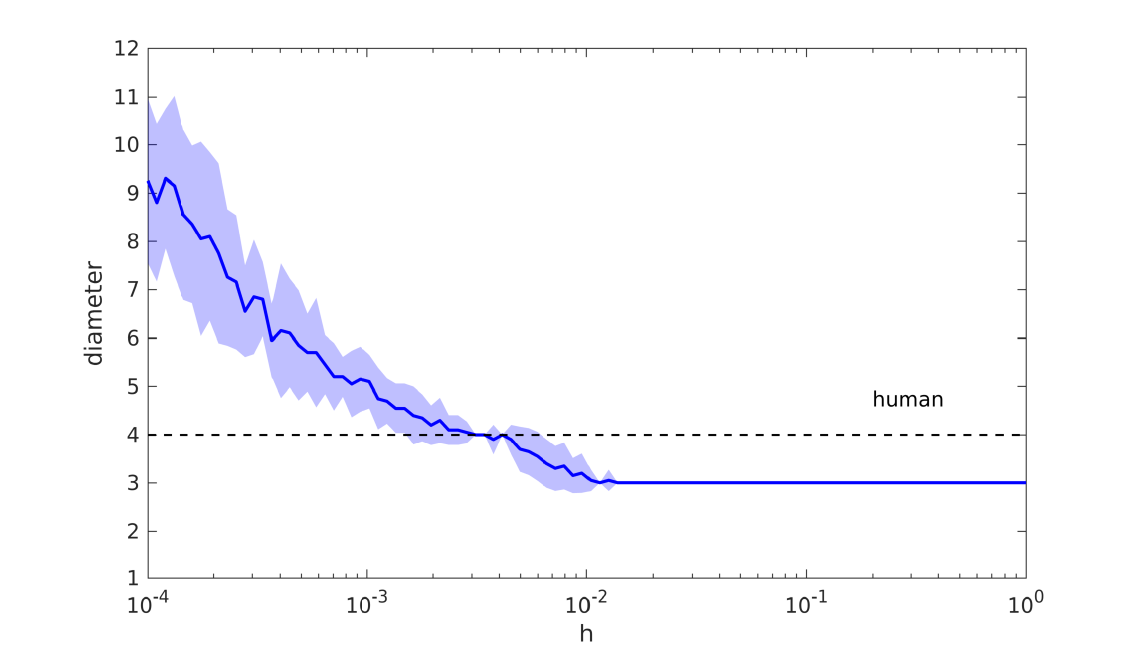}
\caption{
\textbf{Dependence of the graph diameter on the hierarchical
nature of networks with macroscopic attributes identical to a human connectome.}
To show that the path length between any two brain areas can be much shorter than the
number of hierarchical levels into which the network is decomposed, the 
diameter, viz., the longest of all the geodesics between pairs of nodes in a network,
is shown for surrogate networks having the same number of levels (viz., $12$) and
number of nodes occupying each level as the human connectome shown in
Fig.~4 in the main text (see Table S2 for details). 
The diameter is seen to increase as the hierarchical nature of the network becomes more
prominent upon decreasing $h$, viz.,
the ratio of connection densities of non-consecutive
levels ($\rho_{nc}$) to that of consecutive levels ($\rho_{con}$).
The curve is obtained by averaging over
$20$ network realizations for each value of $h$, the  standard deviation being shown as a shaded
region around the curve. For reference, the diameter computed for the empirical
network is indicated by the broken line. }
\label{fig_S7}
\end{figure*}

\begin{figure*}[t!]
\centering
\includegraphics[width=0.8\linewidth]{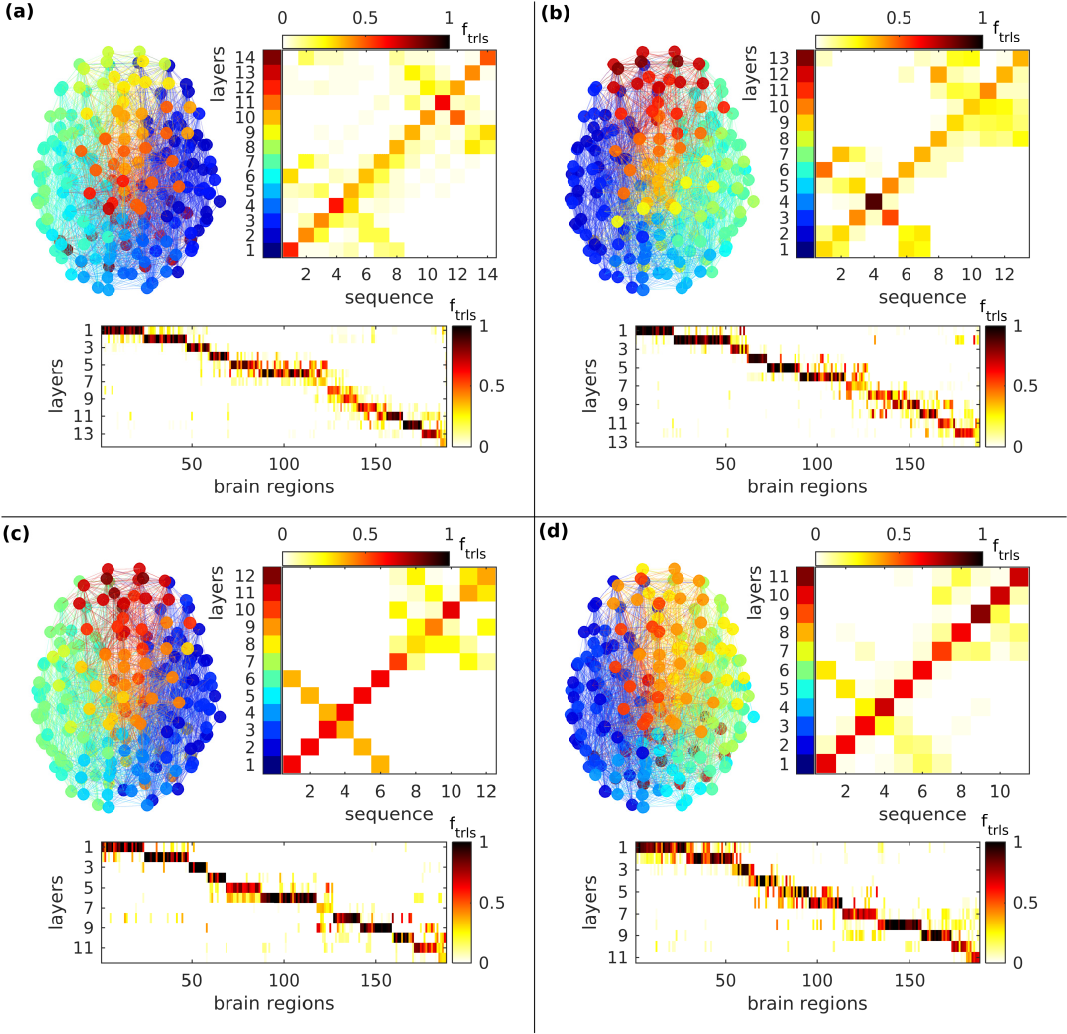}
\caption{
\textbf{Robust hierarchical organization of human connectomes seen across
subjects.}
The panels correspond to four individuals of different
ages, viz., (a) $15$, (b) $27$, (c) $30$ and (d) $42$ years, from the \textit{NKI/Rockland sample}~(see main text for details).
In each panel, the networks of brain areas are shown in horizontal projection (top left) 
with the undirected links representing axonal tracts between the areas. 
The layer in the hierarchy to which an area
(filled circle) belongs is indicated by the corresponding node color (see color key).
Each link between a pair of areas is assigned the color of one of the two nodes it joins.
The robust sequential arrangement of the layers is indicated (top right) by the 
relative frequency $f_{trls}$ of each layer in a reference sequence
(ordered along the ordinate) occurring 
at specific positions (shown along the abscissae) in the hierarchical decomposition obtained in each of $50$ realizations.
The reference sequence is the hierarchical decomposition shown at top left.
The invariance of the hierarchical partitioning of the brain areas identified
across different realizations is quantified (bottom) by the relative frequency $f_{trls}$ with 
which an area occurs at a given layer ordered as per the reference hierarchical arrangement. 
}
\label{fig_S6}
\end{figure*}

\begin{figure*}[t!]
\centering
\includegraphics[width=0.8\linewidth]{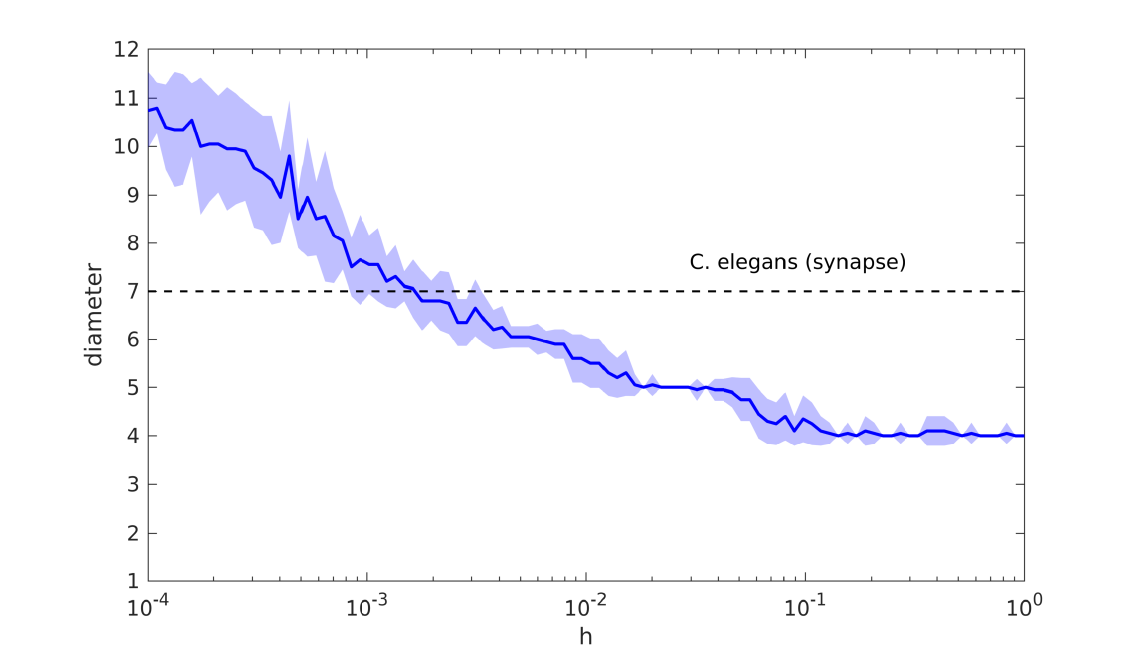}
\caption{
\textbf{Dependence of the graph diameter on the hierarchical
nature of networks with macroscopic attributes identical to the \textit{Caenorhabditis elegans} connectome.}
To show that the path length between any two neurons can be much shorter than the
number of hierarchical levels into which the network is decomposed, the 
diameter, viz., the longest of all the geodesics between pairs of nodes in a network,
is shown for surrogate networks having the same number of levels (viz., $12$) and
number of nodes occupying each level as the \textit{C. elegans} synaptic connectome shown in
Fig.~5 in the main text (see Table S3 for details). 
The diameter is seen to increase as the hierarchical nature of the network becomes more
prominent upon decreasing $h$, viz.,
the ratio of connection densities of non-consecutive
levels ($\rho_{nc}$) to that of consecutive levels ($\rho_{con}$).
The curve is obtained by averaging over
$20$ network realizations for each value of $h$, the  standard deviation being shown as a shaded
region around the curve. For reference, the diameter computed for the empirical
network is indicated by the broken line.}
\label{fig_S9}
\end{figure*}

\begin{figure*}[t!]
\centering
\includegraphics[width=0.49\linewidth]{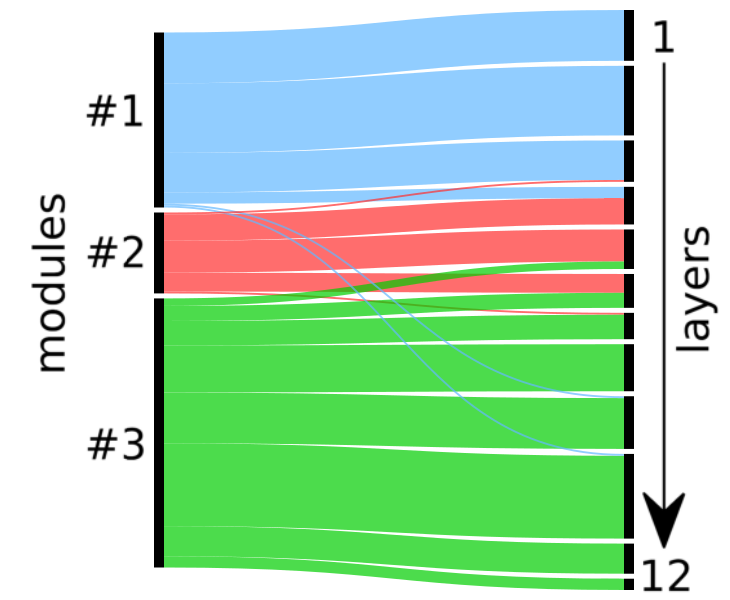}
\includegraphics[width=0.49\linewidth]{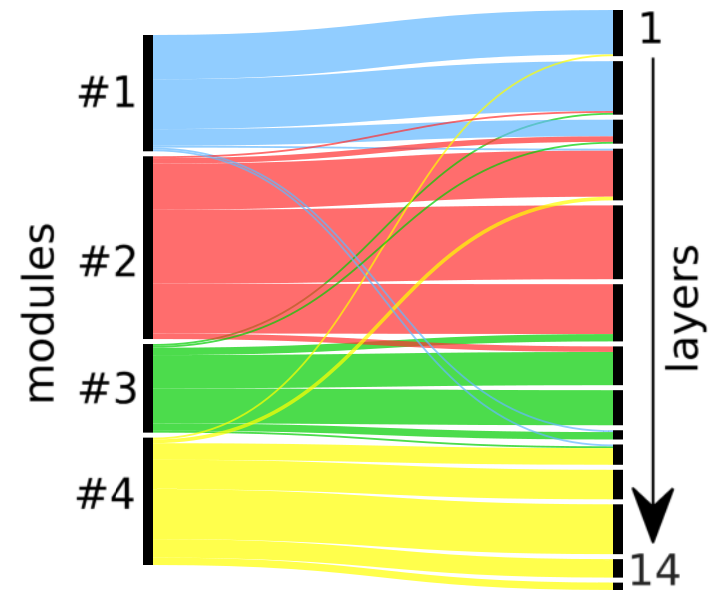}
\caption{
\textbf{Association between modules and hierarchical layers in the
\textit{Caenorhabditis elegans} connectome comprising (left) only synaptic and (right) both synapses and
gap-junctional links between neurons.}
The alluvial diagrams represent the arrangement of the neurons belonging to each
of (left) the three modules identified in the synaptic network and (right)
the four modules determined in the network comprising synapses and gap-junctions,
among the layers in the reference sequences that are shown in Fig.~5 in the main text and Fig.~\ref{fig_S8} in the SI, respectively.}
\label{fig_S11}
\end{figure*}

\begin{figure*}[t!]
\centering
\includegraphics[width=0.8\linewidth]{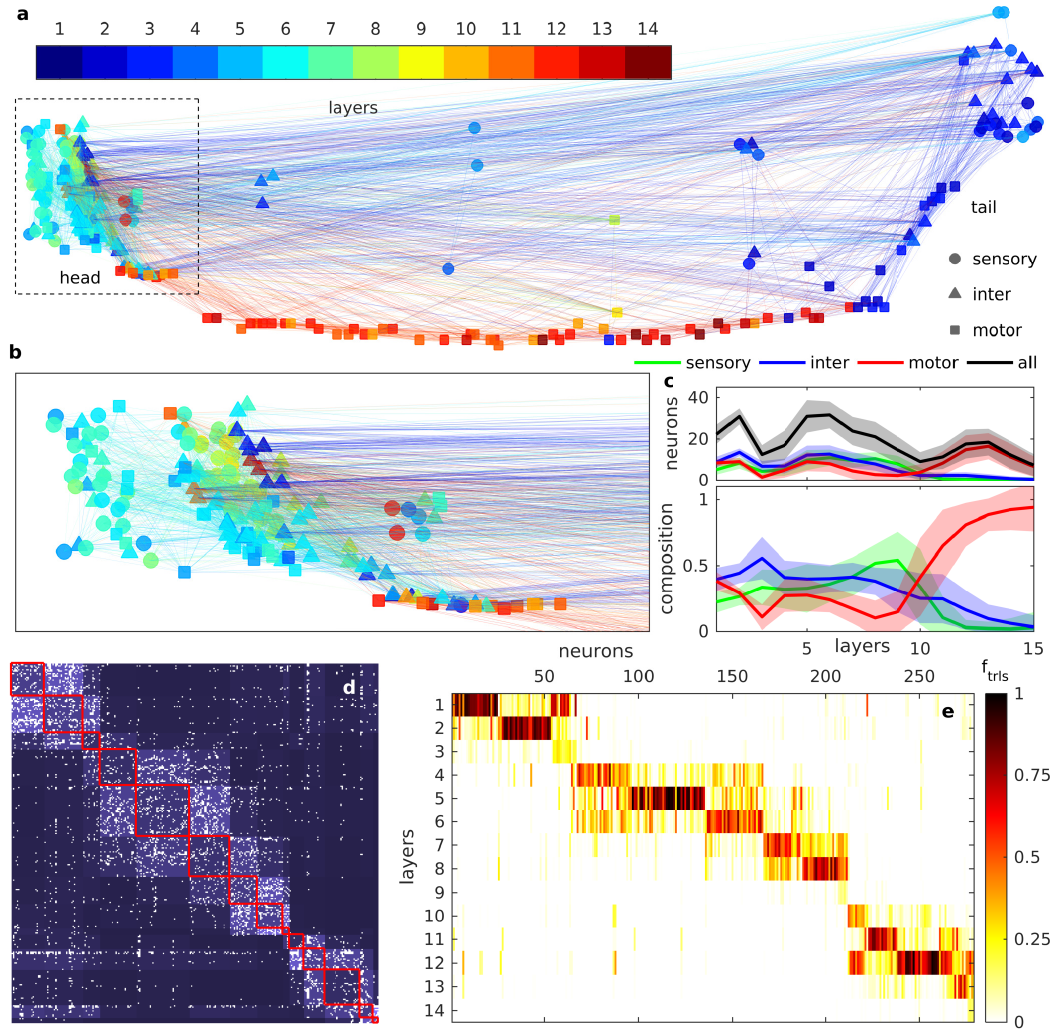}
\caption{
\textbf{The hierarchical structure identified in the \textit{Caenorhabditis elegans}
somatic nervous system network comprising both synaptic and gap junctional connections.}
(a) Spatial representation of the network of synapses and gap-junctions between the $279$ connected neurons that control all activity except pharyngeal movements in the mature
hermaphrodite individuals of the species. The nodes representing the neurons
are arranged according to their position in the worm body along the 
anteroposterior axis, the head and tail being indicated in the figure.
The node color indicates the layer in the hierarchy to which a neuron belongs (see color key),
while the shape indicates whether it is a sensory (circle), motor (square)
or interneuron (triangle).
While each directed synaptic link between a pair of neurons has the same color as the 
source node, an undirected gap-junctional connection is assigned
the color of any one of the two neurons it joins.
To resolve the layered organization of the connections between the densely clustered neurons 
in and around the nerve ring near the head, the area enclosed within the broken lines is shown magnified in panel (b). 
(c) The total number of neurons (black), as well as, the individual functional subtypes, viz.,
sensory (green), motor (red) and interneurons (blue) at each level of the hierarchy 
(upper panel), and the fraction of each subtype in these levels (lower panel).
The solid curve represents the mean while the band represents the
dispersion across $200$ realizations of the hierarchical decomposition of the network. 
(d) Adjacency matrix representation of the \textit{C. elegans} somatic neuronal network,
with nodes (neurons) arranged according to the hierarchical level in which
they occur in the decomposition shown in (a). The existence of a synapse or gap junction 
between a pair of neurons $i,j$ is represented 
by the corresponding entry in the matrix being colored white.
The density of connections between neurons belonging to the same or different hierarchical levels
is indicated by the brightness of the corresponding block. 
(e) The invariance of the hierarchical partitioning of the neurons identified
across different realizations is quantified by the relative frequency $f_{trls}$ with 
which a neuron occurs at a given layer ordered as per the reference hierarchical arrangement 
shown in (d). 
}
\label{fig_S8}
\end{figure*}

\begin{figure*}[t!]
\centering
\includegraphics[width=0.8\linewidth]{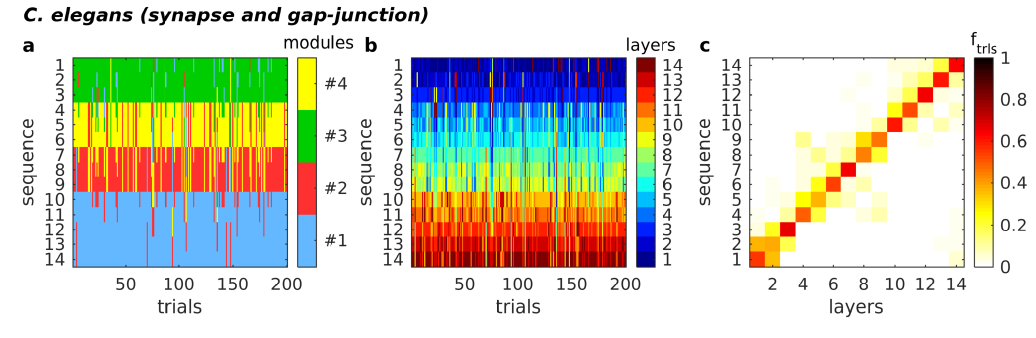}
\caption{
\textbf{Robust sequential arrangement of the modules and hierarchical layers in the
\textit{C. elegans} connectome comprising synaptic, as well as, gap-junctional links between
neurons.}
(a-b) The sequence of layers into which the connectome
is decomposed in each of $200$ 
realizations colored according to (a) the module to which the majority of nodes in each layer belongs and (b) the position they occupy in a reference sequence that
is considered to be the decomposition shown in Fig.~\ref{fig_S8}~(a).
(c)~The relative frequency of occurrence $f_{trls}$, calculated over $200$ realizations, of each layer in the reference sequence (ordered 
along the abscissae) at 
specific positions in the hierarchical decomposition obtained in each trial (shown along the ordinate).
Comparison with the corresponding panels for the synaptic connectome shown 
in Fig.~\ref{fig_S4}~(g-i) indicates that the sequential order of the hierarchical layers is
more consistent across realizations for the network comprising both synapses and gap-junctions.
}
\label{fig_S10}
\end{figure*}

\end{document}